\documentclass[aps,prd,11pt,preprint,tightenlines,unsortedaddress,amsfonts,amssymb,amsmath,byrevtex,showpacs,nofootinbib,onecolumn]{amsart}

\usepackage{latexsym}
\usepackage{geometry}
\usepackage{color}
\usepackage{graphicx}
\usepackage{mathabx}

\usepackage{amssymb,amsmath}
\usepackage{latexsym}

\def\ud{\mathrm{d}}

\begin{document}
\title[]{Primordial gravitational waves \\ in a quantum model of big bounce}
\author[H Bergeron, J-P Gazeau  and P Ma\l kiewicz]{Herv\'e Bergeron, Jean Pierre Gazeau  and Przemys\l aw Ma\l kiewicz}
\address{ISMO, UMR 8214 CNRS, Univ Paris-Sud,  France
} \email{herve.bergeron@u-psud.fr}

\address{ APC, UMR 7164 CNRS, Univ Paris  Diderot, Sorbonne Paris Cit\'e, 75205 Paris, France}\email{gazeau@apc.univ-paris7.fr}

\address{Centro Brasileiro de Pesquisas Fisicas
22290-180 - Rio de Janeiro, RJ, Brazil }\email{gazeau@apc.univ-paris7.fr}

\address{National Centre for Nuclear Research, Ho\.za 69, 00-681 Warsaw, Poland} \email{Przemyslaw.Malkiewicz@ncbj.gov.pl}

\thanks{P.M. was partially supported by MNISW Grant ``Mobilno\'s\'c Plus".}

\date{\today}

\begin{abstract}
We quantise and solve the dynamics of gravitational waves in a quantum Friedmann-Lemaitre-Robertson-Walker spacetime filled with perfect fluid. The classical model is formulated canonically. The Hamiltonian constraint is de-parametrised by setting a fluid variable as the internal clock. The obtained reduced (i.e. physical) phase space is then quantised. {Our quantisation procedure  is implemented in accordance with two different phase space symmetries, namely, the Weyl-Heisenberg symmetry for the perturbation variables, and the affine symmetry for the background variables. As an appealing outcome,} the initial singularity is removed and replaced with a quantum bounce. The quantum model depends on a free parameter that is naturally induced from quantisation and determines the scale of the bounce. We study the dynamics of the quantised gravitational waves across the bounce {through} three different methods (``thin-horizon", analytical and numerical) which give consistent results and we determine the primordial power spectrum for the case of radiation-dominated universe. Next, we use the instantaneous radiation-matter transition transfer function to make approximate predictions for late universe and constrain our model with LIGO and Planck data. We also give an estimate of the quantum uncertainties in the present-day universe. 

\end{abstract}


\maketitle
\tableofcontents

\section{Introduction}
The inflationary scenario assumes a classical homogeneous background spacetime while small inhomogeneous perturbations are quantised \cite{BFM}. {The} quantisation of the metric perturbations is essential for the correct {scenario} predictions of the temperature anisotropies of the cosmic microwave background (CMB). The success of inflation is a strong case in favour of the quantum nature of the gravitational field and makes one feel more confident in wondering about possible quantum effects in the cosmological background evolution. In particular, it encourages alternative cosmological scenarios which, on the one hand, employ the same mechanism of parametric amplification for generating the primordial structures, and, on the other hand, base this mechanism on a quantum background evolution rather than a classical inflationary phase \cite{Brand,cwi}. In the present paper we investigate such a scenario based on the quantum dynamics of tensor perturbations in a quantum Friedmann-Lemaitre-Robertson-Walker (FLRW) universe in which a quantum bounce plays the role of parametric amplifier.

Theories of cosmological perturbations in quantum background spacetimes have been studied before  (see, e.g. \cite{HH,PP1,PP2,AKL}). There are two important aspects of {such} frameworks. Firstly, they always involve a reduced number of quantum variables and they differ in the way the framework is embedded in a fully quantum approach. Secondly, they are based on a certain quantisation of the classically singular cosmological dynamics. To our mind, {a} {\it successful quantisation} should meet at least the two {following} conditions: (i) it should provide a resolution of the cosmological singularity and (ii) it should naturally produce a broad family of quantum models subject to experimental constraints rather than a single model that is completely fixed by a theory and not justified by observational data. In regard to the first aspect our approach bears some similarities to the approach of \cite{AKL} as it starts with a complete quantum formalism which is next reduced. However, the way we reduce the number of quantum variables is different. In regard to the second aspect, the present paper approach is novel and, to our best knowledge, the only one that satisfies the conditions (i) and (ii).

In our approach to quantum dynamics of the cosmological background, the affine coherent states are a basic tool. Let us briefly explain why and how. The homogenous phase space is made from the volume and the expansion rate of the universe and forms the {open} half-plane, $\mathbb{R}^*_+\times\mathbb{R}$. The boundary of this phase space is made of singular states, the volume-vanishing universes. Quantisation is expected to remove the singular states from the dynamics and this can be accomplished {thanks to the appearance of} a quantum repulsive potential that shields the boundary. Such a potential naturally appears in {any} quantisation that respects the specific {affine} symmetry of the homogenous phase space. {Actually, the latter has the structure of the $1$d-affine group}. Notice that the usually employed Weyl-Heisenberg group requires both basic canonical variables to be real and to form a phase space without a  boundary. The {$1$d-}affine group is a minimal group of canonical transformations in the half-plane, which admits two nontrivial unitary irreducible representations (UIR) \cite{isham84, isham_kakas84}.
There is a rich family of affine quantisations. Among them there are such that the basic operators satisfy the canonical commutation rule as in the case of the canonical prescription. On the other hand, the quantum operators that correspond to the Hamiltonian naturally shield the singular states at the boundary.

The affine coherent state quantisation is based on the coherent states that are constructed with a UIR of the {$1$d-}affine group. This quantisation is known to satisfy all the requirements that any quantisation should satisfy such as (i) linearity, (ii) mapping of the constant observable 1 to the identity operator and (iii) possessing the classical limit. The affine coherent state quantisation of the cosmological background dynamics yields a family of singularity-resolving quantum Hamiltonians with free parameters that encode the quantisation ambiguity. As already stated, they all resolve the classical singularity by means of an extra, purely quantum repulsive term that arise due to the boundary. A detailed study of this term in the case of FLRW models can be found in \cite{Bergeron:2013ika}. In this paper we briefly repeat some of the essential steps involved in the discussed quantisation. 

The affine coherent states are also useful for reducing the number of quantum variables of the cosmological background and establishing a framework in which quantum fields propagate in a semiclassical universe. We adopt the idea of Klauder \cite{klauder} and replace the exact quantum motion of the FLRW model by an approximate quantum motion that is confined to the {overcomplete} family of the affine coherent states. This approach leads to a self-consistent derivation of the quantum evolution of gravitational waves coupled to the quantum cosmological background. The evolution of the whole system is given in terms of the coupled Hamilton and Schr\"odinger equations. Solving the equations allows to use the observational bounds from LIGO and Planck on the gravity-wave background amplitude to constrain the free parameters and, indirectly, reveal the scale of the big bounce. It is analogous to the inflationary scenario in which the scale of inflation can be deduced from measurements of the primordial spectra of perturbations and cannot be determined exclusively from the theory. 

Finally, let us briefly describe the classical model that we quantise. We start with the canonical formalism of Arnowitt, Deser and Misner \cite{adm}, which we next truncate at second order for the scalar constraint and at first order for the vector constraints. We neglect the scalar and vector perturbations and focus solely on the tensor perturbations which are at first order gauge-invariant quantities that automatically satisfy the linearised constraints. We employ the Schutz formalism \cite{schutz} for the description of perfect fluid. One of the canonical variables describing the fluid is used to play the role of time and hence, to reduce the initial constrained formalism to an unconstrained formalism with a non-vanishing Hamiltonian. The obtained formalism is then quantised.

The outline of the paper is as follows. In Sec. II we define the classical model, construct the reduced phase space and introduce the truncation of the Hamiltonian to second order in tensor perturbations to the Robertson-Walker geometry. In Sec. III we discuss the {respective quantisations} of  the background geometry and the perturbations. In Sec. IV we introduce a consistent semiclassical approximation to the background evolution and analyse the resulting system of the Hamilton-Schr\"odinger equations. In Sec. V we solve the dynamical equations for the radiation-filled universe. Sec. VI deals with the transfer of the gravitational waves to the late universe. In Sec. VII we use the available observational data to constrain our model. We conclude in Sec. VIII. Appendices \ref{appa}, \ref{appb} and \ref{appC} contain some technical details concerning the affine coherent states and their application to our model. Appendix \ref{appE} contains a list of symbols used throughout the text.

\section{Model of the universe}
In canonical relativity, the spacetime is assumed to admit a foliation $\mathcal{M}=\Sigma\times\mathbf{R}$, where $\Sigma$ is a three-dimensional space-like hyper-surface and $\mathbf{R}$ is the time manifold. Herein, we assume $\Sigma=\mathbb{T}^3$. { The line element of the spacetime reads
\begin{align}
\ud s^2=-N^2\ud t^2+q_{ij}(N^i\ud t+\ud x^i)(N^j\ud t+\ud x^j).
\end{align} }
The Hamiltonian is a sum of four first-class constraints:
\begin{equation}
\mathbf{C}=\int_{\Sigma} N\mathcal{C}_0+N^i\mathcal{C}_i~,
\end{equation}
where the constraints consist of gravitational and matter parts:
\begin{equation}\label{firstcons}
\mathcal{C}_{\nu}=\mathcal{C}_{g,\nu}+\mathcal{C}_{m,\nu}\,.
\end{equation}
In the Arnowitt-Deser-Misner variables \cite{adm}, the gravitational parts read:
\begin{equation}
\mathcal{C}_{g,0}=\sqrt{q}\left(-\frac{1}{2\kappa} {}^{3}R+2\kappa q^{-1}(\pi_a^{~b}\pi_b^{~a}-\frac{1}{2}\pi^2)\right),~~\mathcal{C}_{g,i}=-2D_j\pi^{j}_{~i}\,,
\end{equation}
where $q_{ij}$ are the components of the induced three-metric on $\Sigma$, $\pi^{ij}$ are the components of canonically conjugate momentum, $^3R$ is the Ricci scalar on $\Sigma$, $D_j$ is the covariant derivative on $\Sigma$ and $\kappa=8\pi G$. For the matter part we use the Schultz formalism \cite{schutz}.  For the case of non-rotational perfect fluid, the Hamiltonian constraint is shown to read:
\begin{eqnarray}\label{matcon}
\mathcal{C}_{m,0}=\frac{(p^{\phi})^2}{\frac{1+w}{w}\sqrt{q}K\mu_{h}^{\frac{1-w}{w}}}-\sqrt{q}K\mu_{h}^{\frac{1+w}{w}},~~\mathcal{C}_{m,i}=p^{\phi}\phi_{,i}\, ,
\end{eqnarray}
where $(\phi,p^{\phi})$ is a canonical pair of fluid variables. The meaning of $\mu_{h}$ is the specific enthalpy and its value can be derived from the following relation:
\begin{align}\label{mu}
\mu_{h}^2=\left(\frac{(p^{\phi})^2}{\frac{1+w}{w}\sqrt{q}K\mu_{h}^{\frac{1-w}{w}}}\right)^2-q^{ij}\phi_{,i}\phi_{,j}\, .\end{align}
It is assumed that the fluid satisfies the linear equation of state $p=w\rho$ where $w<1$. The constant $K$ is arbitrary and can be fixed conveniently. 

\subsection{Deparametrisation}
{In the present paper we  study exclusively the tensor perturbations to the FLRW model. The slightly more complicated case of the scalar perturbations will be studied in a future paper. As it is well-known, at linear order the tensor, scalar and vector perturbations to the FLRW metric decouple from each other \cite{bardeen} and thus, their dynamics can be studied separately.} The fluid variables,
\begin{align}
(\phi,~~p^{\phi}),
\end{align}
are a scalar and a densitised scalar, respectively, and can be assumed from the beginning to be {\it homogenous} as the respective perturbations do not couple to tensor perturbations. Hence, the derivatives of the scalar field $\phi_{,i}$ vanish, which simplifies the formulas for the specific enthalpy (\ref{mu}) and the matter constraints (\ref{matcon}):
\begin{eqnarray}\label{simC}
\mu_{h}^{\frac{1}{w}}=\frac{(p^{\phi})^2}{\frac{1+w}{w}\sqrt{q}K},~~~\mathcal{C}_{m,0}=\frac{1}{w}K\sqrt{q}\left(\frac{p^{\phi}}{\frac{1+w}{w}\sqrt{q}K}\right)^{1+w},~~~\mathcal{C}_{m,i}= 0.
\end{eqnarray}
The following canonical transformation,
\begin{align}\label{defS0}
(\phi, p^{\phi}) \mapsto (T,p_T):=\left(\frac{1}{1+w}(p^{\phi})^{\frac{1}{w}}\phi,~(p^{\phi})^{1+w}\right),
\end{align}
makes the scalar matter constraint (\ref{simC}) linear with respect to the fluid momentum, namely
\begin{align}
\mathcal{C}_{m,0}=(\sqrt{q})^{-w}p_T,\end{align}
where we conveniently set $K=w(w+1)^{-\frac{1+w}{w}}$. Therefore, we can solve the scalar constraint, $\mathcal{C}_{g,0}+\mathcal{C}_{m,0}=0$, with respect to the fluid momentum and obtain,
\begin{equation}\label{defS1}
p_T=-(\sqrt{q})^{w}\mathcal{C}_{g,0}.
\end{equation}
Now, the reduced formalism can be achieved by removing the momentum $p_T$ from the phase space and setting the canonically conjugate variable $T$ as the internal clock. This procedure was first proposed by Kuchar in \cite{kuchar} and its later clear expositions can be found in the reviews \cite{Ish, Ku}. Notice that $(\sqrt{q})^{w}\mathcal{C}_{g,0}$ is not a scalar density due to the factor $(\sqrt{q})^{w}$. Nevertheless, in absence of scalar perturbations it can serve, after integration over $\Sigma$, as a non-vanishing physical Hamiltonian,
\begin{equation}\label{H0}
\mathbf{H}=\int_{\Sigma}(\sqrt{q})^{w}\mathcal{C}_{g,0},
\end{equation}
which generates the dynamics in the gravitational variables with respect to $T$ by means of the Poisson bracket,
\begin{align}\label{PB}
\{q_{ab}(x),\pi^{cd}(x')\}=\delta_{(a}^{~c}\delta_{b)}^{~d}\delta^3(x-x').
\end{align}
We emphasise that the internal clock $T$ is not a scalar but a scalar density of weight $\frac{1}{w}$ and as such, in the general case of inhomogenous gravitational field, it defines constant time surfaces ambiguously, i.e. they depend on the employed spatial coordinates. However, since the scalar perturbations are not included, this ambiguity is absent in the present study. 

\subsection{Perturbative expansion}

The reduced Hamiltonian (\ref{H0}) can be expanded to second order around the flat FLRW background. The canonical background variables read:
\begin{equation}\label{defa0}
a^2=\frac{1}{3v_0^{1/3}}\int_{\Sigma}q_{ab}\delta^{ab},~~~p_{a^2}=\frac{1}{v_0^{2/3}}\int_{\Sigma}\pi^{ab}\delta_{ab},~~~\int_{\Sigma}\ud^3x=v_0,
\end{equation}
where $v_0$ is the coordinate volume of $\Sigma=\mathbb{T}^3$. From now on we assume $v_0=1$ {\it dimensionless} and set the three-metric $q_{ab}$ dimensionally 
{\it area} and the conjugate momentum $\pi^{ab}$ dimensionally {\it mass/length}. The canonical perturbation variables are defined as follows:
\begin{align}\label{defP0}
\delta q_{ij}=q_{ij}-a^2\delta_{ij}
,~~\delta\pi^{ij}=\pi^{ij}-\frac{1}{3}p_{a^2}\delta^{ij},
\end{align}
and the Poisson bracket (\ref{PB}) reads now
\begin{align}
\{\delta q_{ab}(x),\delta\pi^{cd}(x')\}=\delta_{(a}^{~c}\delta_{b)}^{~d}\delta^3(x-x'),~~\{a^2,p_{a^2}\}=1.
\end{align}
We restrict the perturbations to the gravitational waves, i.e. to the metric perturbations that are transverse and traceless: 
\begin{align}\label{tensor}\delta q_{ij,i}=0=\delta q_{ii}.\end{align}
We compute the expansion of the Hamiltonian (\ref{H0}) up to second order, 
\begin{align}
\mathbf{H}=\frac{1}{L^{3w-1}}\int_{\Sigma}(\sqrt{q})^{w}\mathcal{C}_{g,0}\simeq \mathbf{H}^{(0)}+\int_{\Sigma}{H}^{(2)}
\end{align}
and obtain:

\begin{align}
\mathbf{H}^{(0)}=&-\frac{\kappa}{3L^{3w-1}}a^{3w+1}p_{a^2}^2\\ \label{fH2}
H^{(2)}=&~\frac{2\kappa}{L^{3w-1}} a^{3w+1}\delta \pi^{ab}\delta\pi_{ab}+\frac{2\kappa}{3L^{3w-1}}a^{3w-1}p_{a^2}\delta\pi^{ab}\delta q_{ab}\\\nonumber
+&\frac{\kappa}{12L^{3w-1}}\frac{5+3w}{3}a^{3w-3}p_{a^2}^2\delta q^{ab}\delta q_{ab}+\frac{1}{8L^{3w-1}\kappa}a^{3w-3}\delta q^{ab,c}\delta q_{ab,c}
\end{align}
where we used $\delta^{ab}$ and $\delta_{ab}$ to respectively rise and lower the indices. Notice that the first-order term vanishes. We introduced the length $L$ in order to have the Hamiltonian which is dimensionally {\it action} and the internal clock which is {\it dimensionless} like the spatial coordinates. This is equivalent to the following choice of the zero-order lapse function,
\begin{align}\label{lapse}
N=\dfrac{a^{3w}}{L^{3w-1}},
\end{align} 
which is dimensionally $length$. { We emphasise that the purpose of introducing the constant length $L$ is to have a dimensionless internal clock, however, the physical predictions do not depend on the particular value of $L$ one fixes for computations.} We simplify the form of the above second-order Hamiltonian by a time-dependent canonical transformation:
\begin{equation}\label{defP1}
(\delta q_{ab},~\delta\pi^{ab})\mapsto \left( w_{ab}:=a^{-2}\delta q_{ab},~ P^{ab}:=a^2\delta \pi^{ab}+\frac{1}{3}p_{a^2}\delta q^{ab}\right),
\end{equation}
where $w_{ab}$ is {\it dimensionless} and $P^{ab}$ is dimensionally {\it action}. The transformation generates extra terms in the Hamiltonian, namely
\begin{align}H^{(2)}\mapsto H^{(2)}+H^{(2)}_{ext}=H^{(2)}{+\frac{2\kappa}{3L^{3w-1}}a^{3w-1}p_{a^2}P^{ab}w_{ab}}-\frac{(w-1)\kappa}{24L^{3w-1}}a^{3w+1}p_{a^2}^2w_{ab} w^{ab}.\end{align} 
Furthermore, we replace the canonical background variables with the ones that we used in our previous paper \cite{Bergeron:2013ika}:
\begin{equation}\label{defa}
(a^2,p_{a^2})\mapsto (q,p):=\left(\gamma a^{\frac{3-3w}{2}},~~\frac{a^{\frac{3w+1}{2}}p_{a^2}}{\sqrt{6}}\right),
\end{equation}
where $\gamma=\frac{4\sqrt{6}}{3(1-w)}$. The final Hamiltonian reads:
\begin{equation}\label{Hfinpos}
\mathbf{H}=-\frac{2\kappa}{L^{3w-1}}p^2+ \int_{\Sigma}\frac{2\kappa}{L^{3w-1}}\frac{\gamma^2}{q^2}P^{ab}P_{ab}+\frac{1}{8\kappa L^{3w-1}}\left(\frac{\gamma}{q}\right)^{\frac{6w+2}{3w-3}}w^{ab,c}w_{ab,c} ~.
\end{equation}
{ The validity of the above result can be directly confirmed by comparing it with Eq. (2.36) of \cite{PP1} after putting $\kappa=3$ (i.e, $\frac{8\pi G}{3}=1$), $k=0$ (i.e. null intrinsic curvature), $N=\frac{a^{3w}}{L^{3w-1}}$ and $q=\gamma a^{\frac{3-3w}{2}}$. One can also verify that our result is in agreement with \cite{langlois,lew}, though one should compare the results of those papers with Eq. (\ref{fH2}) since their authors use $\delta q_{ab}=a^2 w_{ab}$ rather than $w_{ab}$ as a basic variable. The term proportional to $w$ in Eq. (\ref{fH2}) arises due to the second-order perturbation of $(\sqrt{q})^{w}$.}

\subsection{First-order constraints}

The first-order dynamics of the tensor perturbations generated by the Hamiltonian (\ref{Hfinpos}) must be consistent with the linearised vector and scalar constraints (\ref{firstcons}). Let us notice that the vector and scalar constraints at first order must be combinations of terms of a very specific form, namely
\begin{align}
\delta \mathcal{C}_{i}&=c_1w_{jj,i}+c_2w_{ji,i}+c_3P^{jj}_{~~,i}+c_4P^{ji}_{~~,j},\\
\delta \mathcal{C}_{0}&=c_5w_{jj,ii}+c_6w_{ji,ji}+c_7w_{ii}+c_8P^{jj}_{~~,ii}+c_9P^{ji}_{~~,ji}+c_{10}P_{ii},
\end{align}
where $c_i$'s are some zero-order quantities. Since the gravitational waves are traceless and transverse according to the definition (\ref{tensor}), we conclude that all the above terms must vanish,
\begin{align}
\delta \mathcal{C}_{i}\equiv 0,~~\delta \mathcal{C}_{0}\equiv 0,
\end{align}
and hence, the first-order constraints are trivially satisfied. Furthermore, the vanishing constraints do not generate any gauge transformations of the tensor perturbations and thus, the latter are gauge-invariant. This result is in agreement with \cite{langlois,lew}. 

\subsection{Fourier transform}
Recall that we assume $\Sigma=\mathbb{T}^3$. We Fourier-transform the perturbations:
\begin{equation}\label{defFP0}
\widecheck{w}_{ab}(\vec{k})=\int_{\Sigma} {w}_{ab}(\vec{x})\exp(-i\vec{x}\cdot\vec{k})d^3x, ~~~~~~\widecheck{\pi}^{ab}(\vec{k})=\int_{\Sigma} P^{ab}(\vec{x})\exp(-i\vec{x}\cdot\vec{k})d^3x,
\end{equation} 
where $\vec{k}=2\pi\cdot ( n_1, n_2, n_3)$ with $n_i=0,\pm1,\dots$. The reality conditions $\widecheck{w}_{ab}(-\vec{k})=\overline{\widecheck{w}}_{ab}(\vec{k})$ and $\widecheck{\pi}^{ab}(-\vec{k})=\overline{\widecheck{\pi}}^{ab}(\vec{k})$ hold. Furthermore, we express the tensors in the new basis (see \cite{langlois, lew}):
\begin{equation}\label{defFP1}
\widecheck{w}_{ab}=\widecheck{w}_{\pm}A_{ab}^{\pm},~~~~~~\widecheck{\pi}^{ab}=\widecheck{\pi}^{\pm}A_{\pm}^{ab}
\end{equation}
where $A_{ab}^{\pm}:=\frac{k^2}{\sqrt{2}}(v_a w_b\pm v_b w_a)$ and $k=|\vec{k}|$. The vectors $k^{-1}\vec{k}$, $k\vec{v}$ and $k\vec{w}$ form an orthonormal triad. The matrices $A^{ab}_{\pm}$ form the dual basis: $A^{ab}_{\pm}A_{ab}^{\pm}=\delta^{\pm}_{~\pm}$ and $A^{ab}_{\pm}(\vec{k})=A^{ab}_{\pm}(-\vec{k})$. The Hamiltonian in this new basis reads:
\begin{align}\label{hamfinal}
\mathbf{H}=\frac{2\kappa}{L^{3w-1}}p^2-\sum_{\vec{k}} \frac{2\kappa}{L^{3w-1}}\frac{\gamma^2}{q^2}\left(\widecheck{\pi}_+(\vec{k})^2+\widecheck{\pi}_-(\vec{k})^2\right)-\sum_{\vec{k}}\frac{1}{8\kappa L^{3w-1}}k^2\left(\frac{\gamma}{q}\right)^{\frac{6w+2}{3w-3}}\left(\widecheck{w}_+(\vec{k})^2+\widecheck{w}_-(\vec{k})^2\right),
\end{align}
where we changed the time arrow by multiplying the Hamiltonian by $-1$ to be in agreement with the convention of our previous paper \cite{Bergeron:2013ika}.

{
\subsection{Summary}
Let us briefly summarise the steps that we have taken so far. (I) We started from the ADM Hamiltonian formulation of the gravitational field and combined it with the Schutz Hamiltonian formulation of perfect fluid. (II) We solved the respective scalar constraint with respect to the momentum of a fluid while neglecting all the spatial derivatives of the fluid. In this way we arrived at the reduced Hamiltonian formalism with non-vanishing Hamiltonian. (III) Next, we expanded the formalism up to second order in tensor perturbations around the flat FLRW universe. (IV) After making a canonical transformation of both the perturbation and the background variables we arrived at a convenient form of the Hamiltonian, which agrees with similar results present in the literature. (V) Finally, we decomposed the perturbation variables into Fourier modes and obtained the final form of Hamiltonian (\ref{hamfinal}) which completes our formulation of the classical model. Now, we move to the quantum theory.
}

\section{Quantisation}
In the present section we quantise the Hamiltonian (\ref{hamfinal}) which comprises the background variables $(q,p)\in\mathbb{R}_+\times\mathbb{R}$ and the perturbation variables $(\widecheck{w}_{\pm}(\vec{k}),\widecheck{\pi}_{\pm}(\vec{k}))\in\mathbb{R}^2$, where $\vec{k}=(2\pi n_1,2\pi n_2,2\pi n_3)$ with $n_i=0,\pm1,\dots$ For quantisation of the background variables $(q,p)$ we use our previous result obtained with the affine coherent state quantisation of the the Friedmann model \cite{Bergeron:2013ika}. The perturbation variables $(\widecheck{w}_{\pm}(\vec{k}),\widecheck{\pi}_{\pm}(\vec{k}))$ are quantised with the usual canonical prescription. The resultant Hilbert space is
\begin{align}
\mathcal{H}_{tot}=\mathcal{H}_{hom}\otimes\prod_{\vec{k},\pm}\mathcal{H}_{(\vec{k},\pm)}\, ,
\end{align} 
where $\mathcal{H}_{hom}=L^2(\mathbb{R}_+,\ud x)$ and $\mathcal{H}_{(\vec{k},\pm)}=L^2(\mathbb{R},\ud \widecheck{w}_{\pm}(\vec{k}))$. 

\subsection{Background variables}
Let us first recall the {definition of the $1$d-affine group} which is a minimal canonical group of the phase space $(q,p)\in\mathbb{R}_+\times\mathbb{R}$. It is a regular group whose elements can be identified with the phase space points and its {internal} action reads
\begin{align}
(q',p')\circ(q,p)=(q'q,\frac{p}{q'}+p').
\end{align}
{We consider here its} unitary, irreducible and square-integrable representation in $\mathcal{H}_{hom}=L^2(\mathbb{R}_+,\ud x)$, {which is defined as}
\begin{align}
\langle x|U(q,p)|\psi\rangle=\frac{e^{-ipx}}{\sqrt{q}}\psi(x/q),~~\psi(x)=\langle x|\psi\rangle\in L^2(\mathbb{R}_+,\ud x).
\end{align}
{Let us now  define the family of  states},
\begin{align}\label{affcs}
\mathbb{R}_+\times\mathbb{R}\ni (q,p)\mapsto |q,p\rangle:=U(q,p)|\psi\rangle \in\mathcal{H}_{hom},
\end{align} 
where $|\psi\rangle$ is the so called fiducial vector, a normalised fixed vector in Hilbert space such that $\int|\psi(x)|^2\frac{\ud x}{x}<\infty$. {These states  resolve the identity,}
\begin{align}
\int_{\mathbb{R}_+\times\mathbb{R}}\frac{\ud q\ud p}{2\pi c}|q,p\rangle\langle q,p|=\mathbf{1}_{\mathcal{H}}, ~~~ c=\int_{\mathbb{R}_+}|\psi(x)|^2\frac{\ud x}{x} \, , 
\end{align}
{and so will be named \textit{affine coherent states}}. 
{They allow to proceed with the following  affine coherent state quantisation},
\begin{align}\label{qmap}
f(q,p)\mapsto A_f:=\int_{\mathbb{R}_+\times\mathbb{R}}\frac{\ud q\ud p}{2\pi c}f(q,p)|q,p\rangle\langle q,p|\, . 
\end{align}
{This map  is linear, $f=1$ promotes to the identity and semi-bounded functions $f$ promotes to semi-bounded operators $A_f$. Furthermore, and this is essential, this map is covariant with respect to the affine symmetry of the open half-plane, i.e., any point $(q_0,p_0)$ with $q_0 >0$ and $p_0 \in \mathbb{R}$ can be chosen as an origin:
\begin{equation}
\label{covaff} 
U(q_0,p_0) A_f U^{\dag}(q_0,p_0) = A_{\mathfrak{U}(q_0,p_0)f}\, ,
\end{equation}
with
\begin{equation}
\label{covaff2}
 \left(\mathfrak{U}(q_0,p_0)f\right)(q,p)=
f\left((q_0,p_0)^{-1}\circ(q,p)\right)= f\left(\frac{q}{q_0},q_0(p
-p_0) \right)\, .
\end{equation}}

In Appendix \ref{appa} we make a specific choice of a family of the fiducial vectors $|\psi_{\nu}\rangle$ parametrised {by a free parameter $\nu >0$}. Then we prove that the affine coherent state quantisation (\ref{qmap}) of some relevant {basic} observables yields the following result:
\begin{align}
A_{q^{\alpha}}=a_{\alpha}\widehat{q}^{\alpha},~~A_p=\widehat{p},~~A_{p^2}=\widehat{p}^2+\frac{K_1}{\widehat{q}^2},
\end{align}
where $\widehat{q}$ and $\widehat{p}$ are the `position' and `momentum' operators on the half-line, $a_{\alpha}=e^{\frac{\alpha(\alpha-1)}{4\nu}}$ and $K_1=\frac{\nu}{2}$ are positive numbers which depend on the particular choice of the fiducial vector $|\psi_{\nu}\rangle$. The detailed derivation can be found in Appendix \ref{appa}. Keeping $\nu$ unspecified allows for fixing the affine quantisation later, on yet unknown theoretical or observational grounds. Nevertheless, qualitatively the quantisation is unique. In particular, note the repulsive potential $\propto \dfrac{1}{\widehat{q}^{\,2}}$ issued from our quantisation {of the kinetic part of the Hamiltonian, $p^2$}. This term prevents the `particle' from reaching the endpoint $q=0$ by shielding the singularity and reversing the particle's motion. The origin of the extra repulsive potential is due to respecting the affine symmetry. Therefore, when applied to the zero order term of the Hamiltonian (\ref{hamfinal}), the affine CS quantisation removes the singularity and replaces it with a smooth bounce.

\subsection{Perturbation variables}
The canonical perturbation variables, $(\widecheck{w}_{\pm}(\vec{k}),\widecheck{\pi}_{\pm}(\vec{k}))\in\mathbb{R}^2$, are given the {usual} canonical representation on $L^2(\mathbb{R},\ud \widecheck{w}_{\pm}(\vec{k}))$ as operator-valued perturbation variables,
\begin{align}
\widecheck{w}_{\pm}(\vec{k})\mapsto\widehat{w}_{\pm}(\vec{k}),~~\widecheck{\pi}_{\pm}(\vec{k})\mapsto\widehat{\pi}_{\pm}(\vec{k}),~~[\widehat{w}_{\pm}(\vec{k}),\widehat{\pi}_{\pm}(\vec{l})]=i\hbar\delta_{\vec{k}\vec{l}} .
\end{align}
In the next section we will introduce the creation and annihilation operators and work out the dynamics of the perturbation variables in the Heisenberg picture. The quantised Hamiltonian (\ref{hamfinal}) reads:
\begin{equation}\label{qham}
\widehat{\mathbf{H}}=\frac{2\kappa}{L^{3w-1}}\left(\widehat{p}^2+\frac{\hbar^2 K_1}{\widehat{q}^2}\right)-\sum_{k} \frac{\frak{K}_1\gamma^2}{\widehat{q}^2}\left(\frac{2\kappa}{L^{3w-1}}\widehat{\pi}_{\pm,k}^2+\frac{\frak{K}_2}{8\kappa L^{3w-1}}k^2\left(\frac{\gamma}{\widehat{q}}\right)^{\frac{8}{3w-3}}\widehat{w}_{\pm,k}^2\right),
\end{equation}
where $K_1=\frac{\nu}{2}$, $\frak{K}_1=e^{\frac{3}{2\nu}}$ and $\frak{K}_2=\exp\left(\frac{120w-57}{(3w-3)^24\nu}\right)$. We have suppressed the vectorial notation of the Fourier modes $\vec{k}$ and replaced it with the index $k$, which should not lead to confusion. Let us again note that the quantum Hamiltonian (\ref{qham}) includes a new term, the repulsive potential $\dfrac{\hbar^2K_1}{\widehat{q}^2}$, which shields the singularity at $q=0$. {If $K_1\geqslant 3/4$, then the zero order Hamiltonian is essentially self-adjoint and no boundary condition is needed at $q=0$.} The resulting dynamics exhibits a bounce replacing the singularity as we will show shortly. We also note that some coefficients in the quantised Hamiltonian depend on the free parameter $\nu$ defining the family of quantisation maps through the choice of fiducial vector (\ref{fiducial}). The only way to determine the right value of the parameter $\nu$ is to compare the predictions given by our model with the physical Universe. Nevertheless, all values of the parameter $\nu$ produce qualitatively the same Hamiltonian and thus, qualitatively the same dynamics.

\section{Semiclassical approach}
For the purpose of studying the quantum dynamics generated by the Hamiltonian (\ref{qham}) we are going to approximate quantum states of the cosmological background with the affine coherent states. Note that this is a completely different use of the affine CS than previously. We follow the approach introduced by Klauder \cite{klauder}.  Let us describe it briefly. Given a quantum Hamiltonian $\widehat{H}$, the quantum action is defined as follows:
\begin{equation}\label{qaction}
A_Q(\psi):=\int\langle\psi|i\frac{\partial}{\partial T}-\widehat{H}|\psi\rangle~\ud T.
\end{equation}
Variation of (\ref{qaction}) with respect to $\psi$ (or, $\bar{\psi}$) leads to the Schr\"odinger equation:
\begin{equation}
i\frac{\partial}{\partial T}|\psi\rangle=\widehat{H}|\psi\rangle .
\end{equation}
Klauder's idea is to confine the states $|\psi\rangle$ to coherent states, and to introduce the semiclassical action,
\begin{equation}\label{aaction}
A_S(q,p):=\int\langle q,p|i\frac{\partial}{\partial T}-\widehat{H}|q,p\rangle~\ud T.
\end{equation}
When varied with respect to $q$ and $p$ independently, it produces the Hamilton equations,
 \begin{equation}\label{seom}
\frac{\ud q}{\ud T}=\frac{\partial H_S}{\partial p},~~\frac{\ud p}{\ud T}=-\frac{\partial H_S}{\partial q},~~H_S=\langle q,p|\widehat{H}|q,p\rangle .
\end{equation}
The solution to the equations (\ref{seom}) describes at the same time a {\it semiclassical} dynamics in the phase space $(q,p)$ and an approximate quantum dynamics in the Hilbert space,
\begin{align}
\mathbb{R}\ni T\mapsto |q(T),p(T)\rangle \in\mathcal{H}.
\end{align} 

\subsection{Semiclassical Hamiltonian}

In what follows we employ the Klauder idea to the quantum Hamiltonian (\ref{qham}) which comprises both the background variables and the perturbation variables.
We combine the semiclassical approach to the background variables with the full quantum description of the perturbation variables. More precisely, we restrict the total Hilbert space $\mathcal{H}_{tot}$ to the following subset:
\begin{align}\label{defSUBSET}\bigcup_{q,p}\left(|q,p\rangle\otimes\prod_{\vec{k},\pm}\mathcal{H}_{(\vec{k},\pm)}\right)\subset \mathcal{H}_{tot} \, ,\end{align}
where $|q,p\rangle\in\mathcal{H}_{hom}$ are the affine coherent states. Since the use of the affine coherent states is here different {from the quantisation map \eqref{qmap}},  we can define them {through} the formula (\ref{affcs}) based on an independent choice of the fiducial  vector. As before, we will introduce a family of fiducial vectors $|\psi_{\mu}\rangle$ parametrised by $\mu$ rather than a single fiducial vector (see Appendix \ref{appb}). 

The respective semiclassical action reads
\begin{align}\nonumber
A_S(q,p,\phi):=\int\langle \phi, q,p|~i\frac{\partial}{\partial T}-\widehat{\mathbf{H}}~|q,p,\phi\rangle~\ud T\\ \label{SemA}
= \int \left[\dot{q}p+\langle \phi| i\frac{\partial}{\partial T}|\phi\rangle - \langle \phi, q,p|\widehat{\mathbf{H}}|q,p,\phi\rangle~\right]\ud T,
\end{align}
where $|q,p,\phi\rangle=|q,p\rangle\otimes|\phi\rangle\in\mathcal{H}_{hom}\otimes\prod_{\vec{k},\pm}\mathcal{H}_{(\vec{k},\pm)}$ and $\widehat{\mathbf{H}}$ is the Hamiltonian of Eq. (\ref{qham}). Variation of (\ref{SemA}) with respect to $q$, $p$ and $\phi$ leads to the following set of the Hamilton and the Schr\"odinger equations,
\begin{align}\label{semi1}
&\frac{\partial q}{\partial T}=\frac{\partial}{\partial p}\langle \phi, q,p|~\widehat{\mathbf{H}}
~|q,p,\phi\rangle,~\frac{\partial p}{\partial T}=-\frac{\partial}{\partial q}\langle \phi, q,p|~\widehat{\mathbf{H}}~|q,p,\phi\rangle\, ,\\ \label{semi2}
&i\frac{\partial}{\partial T}|\phi\rangle = \langle q,p|~\widehat{\mathbf{H}}~|q,p,\phi\rangle\, .
\end{align}
The expectation value of $\widehat{\mathbf{H}}$ in the coherent states, $\langle q,p|\widehat{\mathbf{H}}|q,p\rangle$, is a $(q,p)$-parametrised operator on $\prod_{\vec{k},\pm}\mathcal{H}_{(\vec{k},\pm)}$, which reads:
\begin{equation}\label{hamsemtotal}
\langle q,p|\widehat{\mathbf{H}}|q,p\rangle=\frac{2\kappa}{L^{3w-1}}\left(p^2+\frac{\hbar^2\frak{L}_1}{q^2}\right)-\sum_{k} \frac{\frak{L}_2\gamma^2}{q^2}\left(\frac{2\kappa}{L^{3w-1}}\widehat{\pi}_{\pm,k}^2+\frac{\frak{L}_3}{8\kappa L^{3w-1}}k^2\left(\frac{\gamma}{q}\right)^{\frac{8}{3w-3}}\widehat{w}_{\pm,k}^2\right),
\end{equation}
where $\frak{L}_1=e^{\frac{3}{2\mu}}(\frac{\nu+\mu}{2}+\frac{1}{4})$, $\frak{L}_2=e^{\frac{3}{2}(\frac{1}{\nu}+\frac{1}{\mu})}$ and $\frak{L}_3=\exp\left[\left(\frac{120w-57}{4(3w-3)^2}\right)\left(\frac{1}{\nu}+\frac{1}{\mu}\right)\right]$ are derived in Appendix \ref{appb}. Notice that now the coefficients depend on both $\nu$ and $\mu$ as the affine coherent states were used twice and each time it was a different family of CS depending on a different parameter: in case of quantisation it was  $\nu$ and in case of computing the expectation values it was $\mu$.

\begin{figure}[!t]
\includegraphics[width=0.4\textwidth]{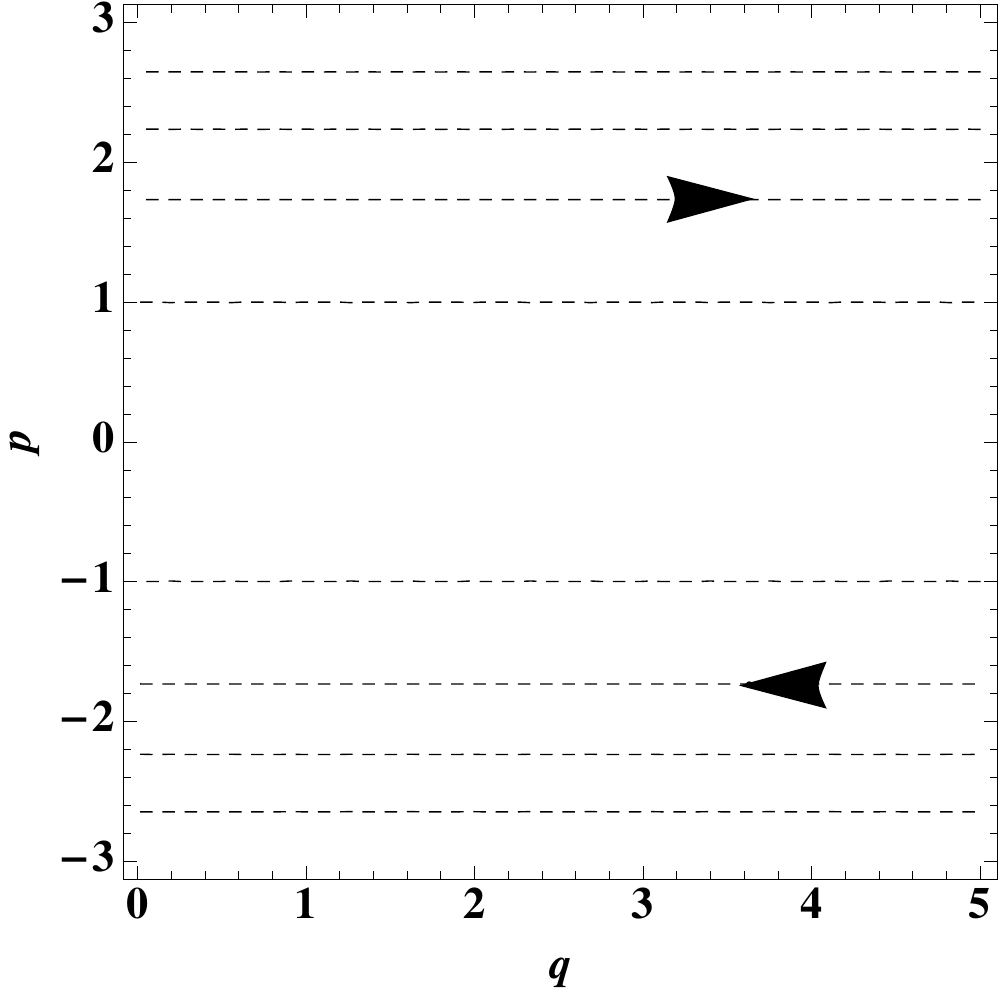}
\hspace{1cm}
\includegraphics[width=0.4\textwidth]{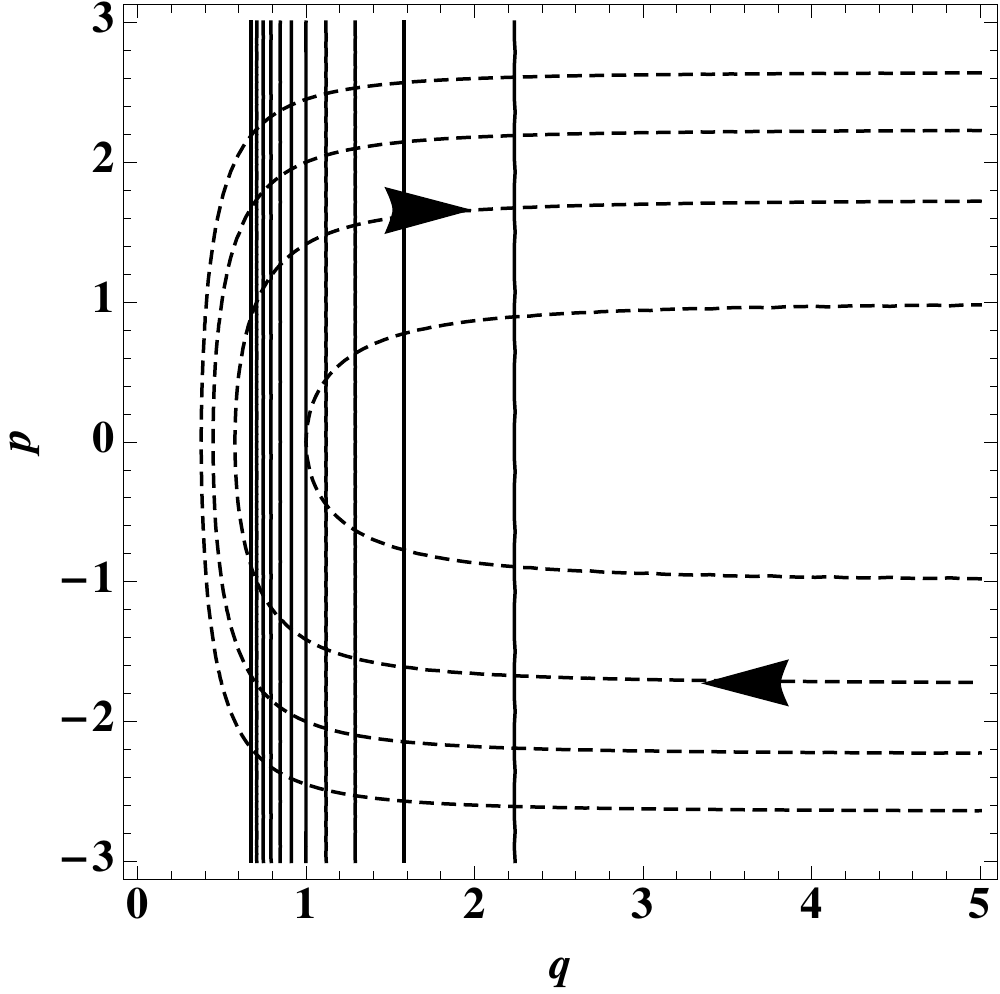}
\caption{{\it On the left:} The classical FLRW phase space trajectories terminate at the $q=0$ singularity. {\it On the right:} The semiclassical FLRW trajectories are generated by a semiclassical Hamiltonian that includes a semiclassical correction, the repulsive potential. The vertical lines are the equipotential lines of the repulsive quantum-induced potential which produces the bounces. In case of radiation ($w=\frac{1}{3}$), $q$ is given in Planck lengths and $p$ is given in Planck masses.}
\label{fig0}
\end{figure}

\subsection{Dynamics of background variables}
Let us first discuss the equations (\ref{semi1}). We note that $\langle q,p,\phi|\widehat{\mathbf{H}}|q,p,\phi\rangle$ comprises zero-order and second-order terms in the perturbation variables and so do the rhs of Eqs (\ref{semi1}). Since we investigate the dynamics up to first order, we neglect the second-order terms in (\ref{semi1}) and obtain the zero-order dynamical equations,
\begin{align}\label{bgeq}
\frac{\partial q}{\partial T}=\frac{2\kappa}{L^{3w-1}}\frac{\partial}{\partial p}\left(p^2+\frac{\hbar^2\frak{L}_1}{q^2}\right),~\frac{\partial p}{\partial T}=-\frac{2\kappa}{L^{3w-1}}\frac{\partial}{\partial q}\left(p^2+\frac{\hbar^2\frak{L}_1}{q^2}\right).\end{align}
We introduce the constant of motion,
\begin{align}\label{defC}
C=\frac{2\kappa}{L^{3w-1}}\left(p^2+\frac{\hbar^2\frak{L}_1}{q^2}\right),
\end{align} 
and find
\begin{equation}\label{heom}
q(T)=\sqrt{\left(4CT^2+\frac{\frak{L}_1\hbar^2}{C}\right)\frac{2\kappa}{L^{3w-1}}},~~~~p(T)=\frac{2CT}{\sqrt{\left(4CT^2+\frac{\frak{L}_1\hbar^2}{C}\right)\frac{2\kappa}{L^{3w-1}}}}.
\end{equation}
Fig. \ref{fig0} illustrates the semiclassical dynamics and compares it with the classical one.

\subsection{Dynamics of perturbation variables}

Let us now discuss the equation (\ref{semi2}). The averaged Hamiltonian $\langle q,p|\widehat{\mathbf{H}}|q,p\rangle$ of Eq. (\ref{semi2}) includes a zero-order term, which is a conserved quantity and is equal to the constant $C$. Thus, it can be neglected without really changing the space of solutions to Eq. (\ref{semi2}). The remaining second-order part of the Hamiltonian, let us denote it by $\widehat{\mathbf{H}}^{(2)}$, reads as an assembly of quantum harmonic oscillators:
\begin{align}\label{h2sem}
\widehat{\mathbf{H}}^{(2)}=\mathcal{S} \sum_{k}\left(\frac{1}{2}\widehat{\pi}_{\pm,k}^2+\frac{1}{2}\omega_{k}^2\widehat{w}_{\pm,k}^2\right),\end{align}
where 
\begin{align}\mathcal{S}= -\frac{4\kappa}{L^{3w-1}}\frac{\frak{L}_2\gamma^2}{q^2(T)}~~\textrm{~~and~~}~~\omega_{k}^2=\frac{\frak{L}_3}{16\kappa^2}k^2\left(\frac{\gamma}{q(T)}\right)^{\frac{8}{3w-3}}.\end{align}
We will employ the Heisenberg picture for studying the dynamics generated by the Hamiltonian (\ref{h2sem}). Computation of $[\widehat{w}_{\pm,k},\widehat{\mathbf{H}}^{(2)}]$ and $[\widehat{\pi}_{\pm,k},\widehat{\mathbf{H}}^{(2)}]$ allows to determine the dynamics of the perturbation operators, $\widehat{w}_{\pm,k}$,
\begin{equation}\label{wave}
\frac{1}{\mathcal{S}}\frac{~d}{dT}\left(\frac{1}{\mathcal{S}}\frac{d\widehat{w}_{\pm,k}}{dT}\right)=-\omega_{k}^2\widehat{w}_{\pm,k}~ .
\end{equation}
It is convenient to work with the cosmological perturbations in conformal time, so we switch to conformal time $a\ud\eta=N\ud T$, where $N=\frac{1}{L^{3w-1}}\left(\frac{q^2}{\gamma^2}\right)^{\frac{3w}{3(1-w)}}$ and $a=\left(\frac{q^2}{\gamma^2}\right)^{\frac{1}{3(1-w)}}$ according to Eqs (\ref{lapse}) and (\ref{defa}), respectively. Hence,
\begin{equation}\label{eta}
\eta=\int\frac{N}{a}\ud T=\int \frac{1}{L^{3w-1}}\left(\frac{q^2}{\gamma^2}\right)^{\frac{3w-1}{3(1-w)}}\ud T\, ,
\end{equation}
and Eq. (\ref{wave}) reads now
\begin{equation}\label{wave2}
\frac{1}{\mathcal{S}_{\eta}}\frac{~d}{d\eta}\left(\frac{1}{\mathcal{S}_{\eta}}\frac{d\widehat{w}_{\pm,k}}{d\eta}\right)=-\omega_{k}^2\widehat{w}_{\pm,k}\,,
\end{equation}
where 
\begin{align}\label{defSeta}\mathcal{S}_{\eta}=\frac{a}{N}\mathcal{S}=-4\kappa\frak{L}_2\left(\frac{q^2}{\gamma^2}\right)^{\frac{2}{3(w-1)}}=-\frac{4\kappa\frak{L}_2}{a^2(\eta)}~.\end{align} 
A more convenient form of Eq. (\ref{wave2}) is obtained after the redefinition of the perturbation operator,
\begin{align}
\widehat{v}_{\pm,k}=a(\eta)\widehat{w}_{\pm,k}.
\end{align} 
Eq. (\ref{wave2}) in terms of the scale factor and $\widehat{v}_{\pm,k}$ reads
\begin{equation}\label{motion}
\widehat{v}_{\pm,k}''+\left(\frak{L}_2^2\frak{L}_3k^2-\frac{a''}{a}\right)\widehat{v}_{\pm,k}=0,
\end{equation}
where $~'~$ denotes differentiation with respect to conformal time $\eta$. This is the usual form of the wave equation which is used for perturbations in classical cosmological backgrounds (see, e.g. \cite{BFM}). However, in our case the scale factor $a(\eta)$ undergoes a bounce instead of an extended period of inflationary expansion. 

{We can see that there is a small semiclassical correction in Eq. (\ref{motion}) to the usual wave equation in the form of the constant factor 
\begin{align}
\frak{L}_2^2\frak{L}_3=\exp\left[\left(\frac{36w^2-32w+17}{12(w-1)^2}\right)\left(\frac{1}{\nu}+\frac{1}{\mu}\right)\right]\geqslant \exp\left[\frac{89}{63}\left(\frac{1}{\nu}+\frac{1}{\mu}\right)\right],
\end{align}
where, in particular, $\frak{L}_2^2\frak{L}_3=\exp\left[\frac{31}{16}\left(\frac{1}{\nu}+\frac{1}{\mu}\right)\right]$ for radiation, $w=\frac{1}{3}$. For the classical background the factor $\mathfrak{L}_2^2 \mathfrak{L}_3$ is equal to unity while in our case it is larger than one and hence the dispersion relation for the gravitational waves is modified. Namely,
\begin{align}\label{speedofgravity}
v_{g}=\frac{\omega}{k}=\sqrt{\frak{L}_2^2\frak{L}_3},
\end{align}
and the speed of gravity waves is larger than unity, i.e. the speed of light. The origin of the factor $\mathfrak{L}_2^2 \mathfrak{L}_3$ is  the fact that we have quantised the cosmological background. On the classical level, the gravitational waves are coupled to the background through some powers of $q$ given in Eq. (\ref{hamfinal}). Upon quantisation the powers of $q$ are promoted to respective operators as in Eq. (\ref{qham}). In the semiclassical approach only the respective expectation values are relevant. However, the expectation value of a given power of $\hat{q}$ is, in general, not equal to the given power of the expectation value of $\hat{q}$, and this naturally leads to the appearance of this sort of factors like $\mathfrak{L}_2^2 \mathfrak{L}_3$. Its value depends, among other things, on the particular quantum state of the background on which one averages. The advantage of our framework is precisely that it includes these kind of quantum effects. The latest measurement of gravitational waves by the LIGO-Virgo collaboration shows that they travel at the speed of light up to very high accuracy \cite{LV}, namely $v_g-1\lesssim 7\cdot 10^{-16}$. This translates into the requirement
\begin{align}
\frac{1}{\nu}+\frac{1}{\mu}\lesssim 10^{-16}~~\textrm{or} ~~~\nu,\mu\gtrsim 10^{16}.
\end{align}
Within our model we can require $\mu$ and $\nu$ to be as large as we wish and hence, make $\mathfrak{L}_2^2 \mathfrak{L}_3$ sufficiently close to unity. {In fact}, as we argue in Sec VII, the allowed values of $\mu$ and $\nu$ must be much larger than $10^{16}$ in order to produce a sufficiently mild bounce that leads to acceptably small amplitude of the primordial gravitational waves.}

\begin{figure}[!t]
\includegraphics[width=0.6\textwidth]{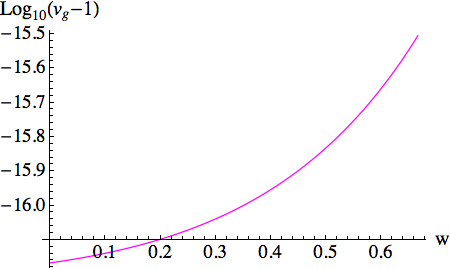}
\caption{The speed of gravity wave $v_g$ ($c=1$) as a function of the fluid used as the internal clock and parametrised by $w=\frac{\mathrm{p}}{\mathrm{\rho}}$. We assumed $\frac{1}{\nu}+\frac{1}{\mu}=10^{-16}$.}
\label{fig_vg}
\end{figure}

{However, notice that the above interpretation of Eq. (\ref{motion}) relies on a very important assumption, that is, that the conformal time used in this equation is identical with the conformal time measured by observers in the present universe. This, however, might not be the case. Notice that Eq. (\ref{eta}) defines the conformal time of the wave equation (\ref{motion}) by means of the transformation of the internal clock $T$ for which the lapse is given in Eq. (\ref{lapse}). Thus, one can argue that even though in the reduced phase space approach the internal clock $T$ remains classical, the corresponding lapse function should be treated as a quantum variable if it depends on variables subsequently quantised like in the present paper. Therefore, it seems justified to modify Eq. (\ref{eta}) as follows
\begin{equation}\label{etaSEM}
\eta=\int\frac{N_{sem}}{a_{sem}}\ud T=\lambda_{sem} \int\frac{N}{a}\ud T\, ,
\end{equation}
where $N_{sem}=\langle q,p|A_{N}|q,p\rangle$, $a_{sem}=\langle q,p|A_{a}| q,p\rangle$ and we find $\lambda_{sem}=e^{\frac{(3w-1)(9w-1)}{18(1-w)^2}\left(\frac{1}{\nu}+\frac{1}{\mu}\right)}$ (recall that $A_O$ stands for a quantum operator corresponding to the observable $O$). For radiation $\lambda_{sem}=1$, however, in general, $\lambda_{sem}\neq 1$. Notice that for different matter contents we use different internal clocks with the different dependence of the lapse function and the scale factor on the basic variable $q$ (see Eqs (\ref{lapse}) and (\ref{defa})). Hence, the time transformation (\ref{eta}) after quantisation and subsequent computation of the expectation values in the affine coherent states will in general become dressed by a different value of $\lambda_{sem}$ as it is apparent from Eq. (\ref{etaSEM}). Subsequently, Eq. (\ref{motion}) becomes modified,
\begin{equation}
\widehat{v}_{\pm,k}''+\left(\frac{\frak{L}_2^2\frak{L}_3}{\lambda_{sem}}k^2-\frac{a''}{a}\right)\widehat{v}_{\pm,k}=0,
\end{equation}
where the factor $\frac{\frak{L}_2^2\frak{L}_3}{\lambda_{sem}}=\exp\left[\frac{27w^2-36w+24\frac{1}{2}}{18(1-w)^2}\left(\frac{1}{\nu}+\frac{1}{\mu}\right)\right]$. Hence, the speed of gravity waves $v_{g}=\sqrt{\frac{\frak{L}_2^2\frak{L}_3}{\lambda_{sem}}}$ depends on the choice of internal clock used for quantisation (see Fig. \ref{fig_vg}). Therefore, one could rise the question of whether there is a preferred internal clock that one should apply in the passing from the initial phase space to the reduced one (irrespectively of the choice of matter content). A most obvious proposal is to choose such an internal clock from which we get $v_{g}=1$. Consequently, for {\it any} choice of clock one should assume such $\lambda_{sem}$ that $v_g=\sqrt{\frac{\frak{L}_2^2\frak{L}_3}{\lambda_{sem}}}=1$ in order to be consistent with the dynamics in the preferred clock.}

{Finally, let us observe that in order to obtain a reliable interpretation of Eq. (\ref{motion}) we first need to derive a quantum dynamics of the Maxwell field in the quantum FLRW background. It is possible that the effective speed of light could turn out to be different from unity or even the same as the effective speed of gravity waves (\ref{speedofgravity}). In the latter case the effective speed of light could be interpreted as the observable value and the observable speed of gravity waves would be simply the same.}

{As some unresolved issues are involved in the interpretation of Eq. (\ref{motion}), we are not presently able to take any definite position in this regard. Therefore, we will simply stick to Eq. (\ref{motion}) as it stands and later prove that $\frak{L}_2^2\frak{L}_3=1$ is a sufficiently good approximation so that we need not worry about the Lorentz-invariance violation by gravitational waves.}

\subsection{Creation and annihilation operators}
We assume the following form of the basic canonical operators:
\begin{equation}\label{defaa}
\widehat{w}_{\pm,k}(\eta)=\frac{1}{\sqrt{2}}\left(\widehat{a}_{\pm,k}w_{\pm,k}^*(\eta)+\widehat{a}_{\pm,k}^{\dagger}w_{\pm,k}(\eta)\right),
~~\widehat{\pi}_{\pm,k}(\eta)=\frac{1}{\sqrt{2}}\left(\widehat{a}_{\pm,k}\frac{1}{\mathcal{S}_{\eta}}\acute{w}_{\pm,k}^*(\eta)+\widehat{a}_{\pm,k}^{\dagger}\frac{1}{\mathcal{S}_{\eta}}\acute{w}_{\pm,k}(\eta)\right),
\end{equation}
where $\widehat{a}_{\pm,k}$ and $\widehat{a}^{\dagger}_{\pm,k}$ are fixed operators and $w_{\pm,k}(\eta)$ are the mode functions which solve Eq. (\ref{wave2}) (or, equivalently $v_{\pm,k}(\eta):=a(\eta)w_{\pm,k}(\eta)$ solve Eq. (\ref{motion})). {Here, ``$\acute{\phantom{A}}$'' stands for derivative with respect to $\eta$.} We demand the canonical commutation relation,
\begin{align}\nonumber
i\hbar&=[\widehat{w}_{\pm,k},\widehat{\pi}_{\pm,k}]\\&=\left[\frac{1}{\sqrt{2}}\left(\widehat{a}_{\pm,k}w_{\pm,k}^*+\widehat{a}_{\pm,k}^{\dagger}w_{\pm,k}\right),
\frac{1}{\sqrt{2}}\left(\widehat{a}_{\pm,k}\frac{1}{\mathcal{S}_{\eta}}\acute{w}_{\pm,k}^*+\widehat{a}_{\pm,k}^{\dagger}\frac{1}{\mathcal{S}_{\eta}}\acute{w}_{\pm,k}\right)\right]\\\nonumber &=
[\widehat{a}_{\pm,k},\widehat{a}_{\pm,k}^{\dagger}]\frac{w_{\pm,k}^*\acute{w}_{\pm,k}-w_{\pm,k}\acute{w}_{\pm,k}^*}{2\mathcal{S}_{\eta}}~.
\end{align}
From Eq. (\ref{wave2}) it follows that
\begin{equation}
\frac{\ud}{\ud\eta}\left(w_{\pm,k}^*\frac{1}{\mathcal{S}_{\eta}}\acute{w}_{\pm,k}-w_{\pm,k}\frac{1}{\mathcal{S}_{\eta}}\acute{w}_{\pm,k}^*\right)=0~,
\end{equation}
and we choose to fix 
$$w_{\pm,k}^*\frac{1}{\mathcal{S}_{\eta}}\acute{w}_{\pm,k}-w_{\pm,k}\frac{1}{\mathcal{S}_{\eta}}\acute{w}_{\pm,k}^*=2i\hbar~.$$
Thus, the operators $\widehat{a}_{\pm,k}$ and $\widehat{a}^{\dagger}_{\pm,k}$ are the annihilation
and creation operators, respectively. Time evolution is completely encoded into the mode functions $w_{\pm,k}(\eta)$.

After switching to conformal time the Hamiltonian (\ref{h2sem}) reads
\begin{equation}\label{resham}
\begin{split}
\widehat{\mathbf{H}}^{(2)}=\mathcal{S}_{\eta}\sum_k\left(\frac{1}{2}\widehat{\pi}_{\pm,k}^2+\frac{1}{2}\omega_{k}^2\widehat{w}_{\pm,k}^2\right)=\mathcal{S}_{\eta}\sum_k\frac{\widehat{a}_{\pm,k}^2}{4}
\left(\frac{(\acute{w}_{\pm,k}^*)^2}{\mathcal{S}_{\eta}^2}+\omega_{k}^2(w_{\pm,k}^*)^2\right)\\+\mathcal{S}_{\eta}\sum_k\frac{(\widehat{a}_{\pm,k}^{\dagger})^2}{4}\left(\frac{\acute{w}_{\pm,k}^2}{\mathcal{S}_{\eta}^2}+\omega_k^2w_{\pm,k}^2\right)+\mathcal{S}_{\eta}\sum_k\frac{2\widehat{a}_{\pm,k}^{\dagger}\widehat{a}_{\pm,k}+1}{4}\left(\frac{|\acute{w}_{\pm,k}|^2}{\mathcal{S}_{\eta}^2}+\omega_k^2|w_{\pm,k}|^2\right),
\end{split}
\end{equation}
(it is a rescaled version of the Hamiltonian (\ref{h2sem}) but we keep the same notation). We require the vacuum state, i.e. the state $|0\rangle$ such that $\widehat{a}_{\pm,k}|0\rangle=0$ for all $k$, to minimise the energy (\ref{resham}) at some initial moment $\eta_0$. It follows that (we put $\hbar=1$)
\begin{equation}\label{ini}
w_{\pm,k}(\eta_0)=\frac{1}{\sqrt{\omega_{k}(\eta_0)}},~~\frac{\acute{w}_{\pm,k}(\eta_0)}{\mathcal{S}_{\eta}(\eta_0)}=-i{\sqrt{\omega_{k}(\eta_0)}}
\end{equation}
or, equivalently
\begin{equation}\label{ini2}
v_{\pm,k}(\eta_0)=\frac{a(\eta_0)}{\sqrt{\omega_{k}(\eta_0)}},~~\frac{\acute{v}_{\pm,k}(\eta_0)}{\mathcal{S}_{\eta}(\eta_0)}=-ia(\eta_0){\sqrt{\omega_{k}(\eta_0)}}\left(1-\frac{i\acute{a}(\eta_0)}{a(\eta_0)\mathcal{S}_{\eta}(\eta_0)\omega_{k}(\eta_0)}\right)~,
\end{equation}
and the Hamiltonian (\ref{resham}) at $\eta_0$ reads
\begin{equation}
\widehat{\mathbf{H}}^{(2)}(\eta_0)=\mathcal{S}_{\eta}\sum_k\left(a_{\pm,k}^{\dagger}a_{\pm,k}+\frac{1}{2}\right)\omega_{k}(\eta_0)~.
\end{equation}
We assume that away from the bounce, for $\eta_0\rightarrow \pm\infty$, the scale factor $a$ is slowly-varying, i.e. $\frac{\acute{a}(\eta_0)}{a(\eta_0)}\ll \frak{L}_2\sqrt{\frak{L}_3}k$, and that the initial condition (\ref{ini2}) can be approximated as:
\begin{equation}\label{iniapp}
v_{\pm,k}(\eta_0)=\frac{a(\eta_0)}{\sqrt{\omega_{k}(\eta_0)}},~~\frac{\acute{v}_{\pm,k}(\eta_0)}{\mathcal{S}_{\eta}(\eta_0)}=ia(\eta_0){\sqrt{\omega_{k}(\eta_0)}}~.
\end{equation}
The average number of particles with momentum $k$ in the state $|0\rangle$ at some later time $\eta_1$ is given by the following formula:
\begin{equation}\label{garvprod}
\langle 0|N_k(\eta_1)|0\rangle=\frac{\langle 0|\widehat{\mathbf{H}}^{(2)}_{k}(\eta_1)|0\rangle}{\mathcal{S}_{\eta}(\eta_1)\omega_k(\eta_1)}-\frac{1}{2}=\frac{\frac{|\acute{w}_{\pm,k}(\eta_1)|^2}{\mathcal{S}_{\eta}^2(\eta_1)}+\omega_{k}^2(\eta_1)|w_{\pm,k}(\eta_1)|^2}{4\omega_{k}(\eta_1)}-\frac{1}{2}~,
\end{equation}
or, in terms of $v_{\pm,k}$,
\begin{equation}\label{Nv}
\langle 0|N_k(\eta_1)|0\rangle= \frac{\frac{|\acute{v}_{\pm,k}(\eta_1)|^2}{\mathcal{S}_{\eta}^2(\eta_1)}+\omega_{k}^2(\eta_1)|v_{\pm,k}(\eta_1)|^2}{4a^2({\eta_1})\omega_{k}(\eta_1)}-\frac{1}{2}~,
\end{equation}
where we shall assume $\frac{\acute{a}(\eta_1)}{a(\eta_1)}\ll \frak{L}_2\sqrt{\frak{L}_3}k$, which holds far away from the bounce.

\section{Case study: radiation-filled universe}
In the present section we assume that the universe is filled with radiation, which appears reasonable for studying its dynamics in the vicinity of a cosmological bounce. For radiation ($w=\frac{1}{3}$), $\mathcal{S}=\mathcal{S}_{\eta}$ in accordance with Eqs (\ref{eta},\ref{defSeta}) and we set $T=\eta$. Then, {from Eq. \eqref{heom}},
\begin{align}\label{aR}
a(\eta)\gamma=q(\eta)=\sqrt{2\kappa\left(4C\eta^2+\frac{\frak{L}_1\hbar^2}{C}\right)}.\end{align}
We can view Eq. (\ref{motion}) as the stationary Schr\"odinger equation for a particle on the real line with the potential barrier $V(\eta)$ and of momentum $\frak{L}_2\sqrt{\frak{L}_3}k$, where:
\begin{equation}
V(\eta)=\frac{a''}{a}=4\frak{L}_1\hbar^2\left(4C\eta^2+\frac{\frak{L}_1\hbar^2}{C}\right)^{-2}
\end{equation}
and $\textrm{max} (V)=\dfrac{4C^2}{\frak{L}_1\hbar^2}$. It is useful to introduce a new notation:
\begin{equation}\label{defk}
\mathbf{k}=\frak{L}_2\sqrt{\frak{L}_3}k,~~~~\mathbf{k}_*=\frac{2C}{\sqrt{\frak{L}_1}\hbar}=\sqrt{\textrm{max} (V)},
\end{equation}
so the equation of motion (\ref{motion}) reads now:
{\begin{align}\label{eom}
&v_{\pm,\mathbf{k}}''+\left[\mathbf{k}^2-V(\eta)\right]v_{\pm,\mathbf{k}}=0\, , \\
\label{eomV} &\mbox{with}\  V(\eta)=\left(\mathbf{k}_*\eta^2+\mathbf{k}_*^{-1}\right)^{-2}\, ,
\end{align}
and for which the initial condition  \eqref{ini2} becomes}
\begin{equation}
v_{\pm,\mathbf{k}}(\eta_0)=\frac{\sqrt{4\kappa\frak{L}_2}}{\sqrt{\mathbf{k}}},~~\acute{v}_{\pm,\mathbf{k}}(\eta_0)\approx i\sqrt{4\kappa\frak{L}_2}\sqrt{\mathbf{k}}.
\end{equation}
The potential $V(\eta)$ originates from the quantum repulsive force issued from our quantisation of the background geometry, which is responsible for the smooth bounce, {already noticed with Eq.\eqref{heom}}. The potential is rather steep. For $C=2$, $\hbar=1$, $\frak{L}_1=9$, it is plotted in Fig. (\ref{fig1}).

\begin{figure}[!t]
\includegraphics[width=0.6\textwidth]{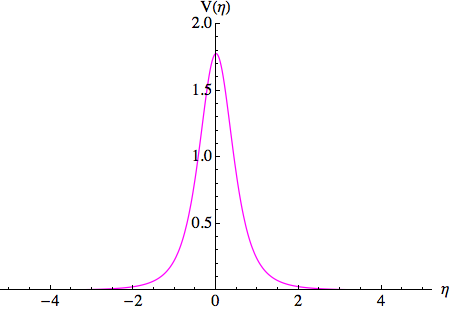}
\caption{Potential barrier for gravitational waves due to the bounce. The problem of evolution of the mode functions is identical to the scattering problem for the above potential and given by the stationary Schr\"odinger equation: $-\frac{\partial^2 \psi}{\partial\eta^2}+V(\eta)\psi=\mathbf{k}^2\psi$. We set $C=2$, $\hbar=1$, $\frak{L}_1=9$.}
\label{fig1}
\end{figure}
Only the waves of small enough momentum $\mathbf{k}<\mathbf{k}_{*}$ can be significantly affected by the barrier. Any solution to the Schr\"odinger equation (\ref{eom}) that comprises an ingoing wave to the left of the potential, will comprise both ingoing and outgoing waves to the right of the potential. Within the formalism developed in the previous section, this phenomenon is associated with the amplification of the amplitude of the perturbation variables, $\widehat{w}_{\pm,k}$. In the Schr\"odinger picture, this is understood in terms of the Bogoliubov transformation of the vacuum state and, consequently, re-interpretation of the initial vacuum state in terms of an excited state full of particles, the gravitons \cite{Mukhanov}. In the spacetime picture, the inverse of potential $V$ describes the size of the horizon which shrinks before the bounce and expands afterwards. Sufficiently long modes of the perturbation exit the horizon before the bounce and re-enter it some time after the bounce. When a wave is outside the horizon, amplification of its amplitude takes place (see e.g. \cite{Brand}). In what follows we find the solution to Eq. (\ref{eom}). 

\subsection{Thin-horizon approximation}
A useful approach to the problem is given by the following thin-horizon approximation. The potential {\eqref{eomV}} fixes the size of the {\it horizon}. The sub-horizon modes are those with
\begin{equation}
\mathbf{k}\gg \frac{1}{\mathbf{k}_*\eta^2+\mathbf{k}_*^{-1}}~,
\end{equation}
whereas the super-horizon modes are those with
\begin{equation}
\mathbf{k}\ll \frac{1}{\mathbf{k}_*\eta^2+\mathbf{k}_*^{-1}}~.\end{equation}
For each mode function, the {\it horizon crossing} happens at
\begin{equation}\label{hcross}
\eta_{\pm}=\pm \frac{1}{\mathbf{k}_*}\sqrt{\frac{\mathbf{k}_{*}}{\mathbf{k}}-1}~.
\end{equation}
Modes with $\mathbf{k}>\mathbf{k}_{*}$ never exit the horizon, and consequently they are not amplified during the bounce. The remaining modes exit the horizon in the contracting phase and enter it again in the expanding phase. In what follows we focus on their dynamics.

\subsubsection{Solutions}
The modes start as sub-horizon modes and their dynamics for $\eta<\eta_-$ in agreement with the initial state (\ref{iniapp}) read:
\begin{equation}\label{ap1}
v_{\mathbf{k}}\approx \frac{\sqrt{4\kappa\frak{L}_2}}{\sqrt{\mathbf{k}}}e^{i\mathbf{k}\eta}~,
\end{equation}
When they become super-horizon for $\eta_-<\eta<\eta_+$, they evolve as
\begin{equation}\label{ap2}
v_{\mathbf{k}}=\sqrt{4\kappa \frak{L}_2}\sqrt{\mathbf{k}_*\eta^2+\mathbf{k}_*^{-1}}\left(A_{\mathbf{k}}+B_{\mathbf{k}}\arctan\left(\mathbf{k}_*\eta\right)\right)\propto a\left(A_{\mathbf{k}} +B_{\mathbf{k}}\int\frac{d\eta}{a^2}\right)~,
\end{equation}
where $A_{\mathbf{k}}$ and $B_{\mathbf{k}}$ are determined by sewing the two above solutions (\ref{ap1}) and (\ref{ap2}) at the time of horizon crossings $\eta_-$. The solution for a mode which has re-entered the horizon ($\eta>\eta_+$) reads
\begin{equation}\label{fvk1}
v_{\mathbf{k}}\approx \frac{\sqrt{4\kappa\frak{L}_2}}{\sqrt{\mathbf{k}}}\left(C_{\mathbf{k}}e^{i{\mathbf{k}}\eta}+D_{\mathbf{k}}e^{-i{\mathbf{k}}\eta}\right)
\end{equation}
where $C_{\mathbf{k}}$ and $D_{\mathbf{k}}$ are determined from sewing the solution (\ref{ap2}) with (\ref{fvk1}) at the horizon crossing $\eta_+$.
We find that:
\begin{eqnarray}\label{defz}\begin{split}
A_{\mathbf{k}}&=\left[1+\left(i+\sqrt{z-1}\right)\arctan\left(\sqrt{z-1}\right)\right]e^{-iz^{-1}\sqrt{z-1}},\\
B_{\mathbf{k}}&=\left[i+\sqrt{z-1}\right]e^{-iz^{-1}\sqrt{z-1}},\\
C_{\mathbf{k}}&=-ie^{-i2z^{-1}\sqrt{z-1}}\left(i+\sqrt{z-1}\right)\left[1+\left(i+\sqrt{z-1}\right)\arctan\left(\sqrt{z-1}\right)\right],\\
D_{\mathbf{k}}&=i\left(\sqrt{z-1}+z\arctan\left(\sqrt{z-1}\right)\right),\end{split}
\end{eqnarray}
where $z=\frac{\mathbf{k}_{*}}{\mathbf{k}}$.

\subsection{Analytical approximation}

\begin{figure}[t]
\includegraphics[width=0.6\textwidth]{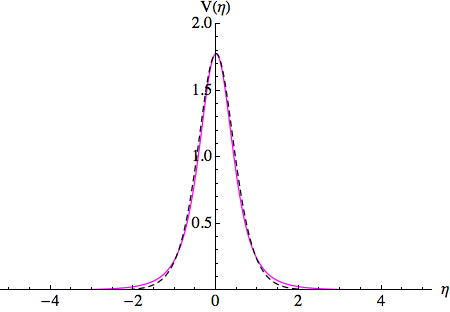}
\caption{The fitting potential (given by the dashed black line) and and the exact potential (given by the colour solid line) are shown to be very similar. We set $C=2$, $\hbar=1$, $\frak{L}_1=9$.}
\label{fig1}
\end{figure}

Let us rewrite Eq. (\ref{eom}) in terms of a new  independent variable $\eta\mapsto u =\mathbf{k}_{*}\eta$,
\begin{equation}\label{neom}
-v_{\mathbf{k},uu}+\frac{1}{(u^2+1)^2}v_{\mathbf{k}}=z^{-2}v_{\mathbf{k}}~,
\end{equation}
where $z=\frac{\mathbf{k}_{*}}{\mathbf{k}}$. This equation is not explicitly solvable. However, we can approximate the potential by another one which leads to an explicit solution. Namely, we notice that numerically we have
\begin{equation}
\frac{1}{(u^2+1)^2} \simeq \frac{1}{\cosh^2(4 u /\pi)}\,,
\end{equation}
where by imposing $\int_{\mathbb{R}} \frac{{\rm d} u}{(u^2+1)^2} = \int_{\mathbb{R}} \frac{{\rm d} u}{\cosh^2(\alpha u)}$ the coefficient $\alpha  = 4/\pi$ is obtained. The two potentials are plotted in Fig. \ref{fig1}. Thus, Eq. \eqref{neom} can be approximated by
\begin{equation}
-v_{\mathbf{k},uu}+\frac{1}{\cosh^2(4 u /\pi)}v_{\mathbf{k}}=z^{-2}v_{\mathbf{k}}
\,.
\end{equation}
Upon rescaling the independent variable again, $u \mapsto \xi = 4u/ \pi$, we obtain
\begin{equation}
\label{schro3}
-v_{\mathbf{k},\xi\xi} + \frac{(\frac{\pi}{4})^2}{\cosh^2 \xi} v_{\mathbf{k}} = {l}_{\mathbf{k}}^2 v_{\mathbf{k}} \,,
\end{equation}
where ${l}_{\mathbf{k}} = z^{-1}\pi/4$. 

\subsubsection{Solutions}

We define the coefficient $s = \frac{1}{2} \left(-1+\sqrt{1-\frac{\pi^2}{4}} \right)$. The solution of Eq. \eqref{schro3} compatible with the {requested} asymptotic behaviour reads
\begin{equation}\label{DEFs}
v_{\mathbf{k}} = \frac{\sqrt{4\kappa\frak{L}_2}}{\sqrt{\mathbf{k}}}\, 2^{-i{l}_{\mathbf{k}}}(1-(\tanh \xi)^2)^{i{l}_{\mathbf{k}}/2} \, _2F_1\left(i{l}_{\mathbf{k}}-s,i{l}_{\mathbf{k}}+s+1;1+i{l}_{\mathbf{k}};\frac{1+\tanh \xi}{2} \right)~.
\end{equation}
From the properties of hypergeometric functions, we find the asymptotic behaviour,
\begin{equation}\label{fvk2}
\left.
\begin{array}{c}
v_{\mathbf{k}} \simeq_{\xi \to -\infty} \, \frac{\sqrt{4\kappa\frak{L}_2}}{\sqrt{\mathbf{k}}}e^{i {l}_{\mathbf{k}} \xi}~,\\
v_{\mathbf{k}} \simeq_{\xi \to +\infty} \frac{\sqrt{4\kappa\frak{L}_2}}{\sqrt{\mathbf{k}}}\left(C_{\mathbf{k}}\, e^{i {l}_{\mathbf{k}} \xi} + D_{\mathbf{k}} \, e^{-i {l}_{\mathbf{k}} \xi}\right)~,
\end{array}\right.
\end{equation}
where
\begin{equation}
\left.
\begin{array}{c}
C_{\mathbf{k}}= \dfrac{\Gamma(1+i{l}_{\mathbf{k}}) \Gamma(i{l}_{\mathbf{k}})}{\Gamma(i{l}_{\mathbf{k}}-s) \Gamma(i{l}_{\mathbf{k}}+s+1)}~, \\
D_{\mathbf{k}} = \dfrac{\Gamma(1+i{l}_{\mathbf{k}}) \Gamma(-i{l}_{\mathbf{k}})}{\Gamma(1+s) \Gamma(-s)}~.
\end{array}\right.
\end{equation}

\subsection{Amplification of the perturbation amplitude}

The amplitude of each of the two polarisation modes  {$\widehat{w}_{ab}:= \widehat{w}_{\pm}A^{\pm}_{ab}$} of the primordial tensor perturbation can be deduced from the equal-time correlation function:
\begin{eqnarray}\label{ampl}
\begin{split}
&\frac{1}{2}\langle 0|\widehat{w}_{ab}(\vec{x})\widehat{w}^{ab}(\vec{y})|0\rangle=\frac{1}{2a^2}\sum_{\vec{k}}e^{i\vec{k} (\vec{x}-\vec{y})}|v_{\mathbf{k}}|^2\approx \frac{1}{2a^2}\int e^{i\vec{k} (\vec{x}-\vec{y})}|v_{\mathbf{k}}|^2 \frac{\ud^3k}{(2\pi)^3}\\
&=\frac{1}{(2\pi)^2a^2}\int |v_{\mathbf{k}}|^2\frac{\sin(k|\vec{x}-\vec{y}|)}{k|\vec{x}-\vec{y}|}k^2\ud k\sim {k_{m}^3\frac{|v_{\mathbf{k}_{m}}|^2}{a^2}~~\textrm{for}~~k_{m}\sim |\vec{x}-\vec{y}|^{-1}}\,,
\end{split}
\end{eqnarray}
where we assume that the interesting coordinate distances $|\vec{x}-\vec{y}|\ll 1$ are small in comparison to the size of the universe (or, $k\gg 1$) so that the sum can be approximated by the integral and the isotropy $v_{\vec{\mathbf{k}}}=v_{\mathbf{k}}$ is used. {In the sequel, we write $k_{m} = k$ for the sake of simplicity.} Following \cite{Mukhanov}, we define the spectrum of amplitude of quantum fluctuations of the gravitational waves as (per each polarisation mode):
\begin{equation}\label{spdef}
\delta_{\widehat{w}}(k)=\frac{|v_{\mathbf{k}}|}{2\pi a}k^{\frac{3}{2}}~.
\end{equation}
Taking into account the asymptotic behaviour of the mode functions given in Eqs (\ref{fvk1}) and (\ref{fvk2}), the amplitude spectrum reads (we restore $\hbar$):
\begin{equation}
\delta_{\widehat{w}}(k)=\frac{1}{\frak{L}_2\frak{L}_3^{3/4}}~\cdot\frac{\sqrt{|C_{\mathbf{k}}|^2(z)+|D_{\mathbf{k}}|^2(z)}}{\pi z}~\cdot\left[\sqrt{\hbar\kappa}\frac{~\mathbf{k}_*}{~V^{1/3}}\right]\,,
\end{equation}
where we have time-averaged the oscillatory part and where $z=\frac{\mathbf{k}_*}{\mathbf{k}}$. Initially, for $\eta=\eta_0$ we have the amplitude of the vacuum fluctuations,
\begin{equation}
\delta_{\widehat{w}}(k)=\frac{1}{\frak{L}_2\frak{L}_3^{3/4}}~\cdot\frac{1}{\pi z}~\cdot\left[\sqrt{\hbar\kappa}\frac{~\mathbf{k}_*}{~V^{1/3}}\right],
\end{equation}
where the coefficient $\frac{1}{\frak{L}_2\frak{L}_3^{3/4}}$ is very close to $1$ as we have pointed out below Eq. (\ref{motion}) (see Sec VII for a discussion) and thus, in what follows we set $\frac{1}{\frak{L}_2\frak{L}_3^{3/4}}=1$.

\begin{figure}[!t]
\includegraphics[width=0.45\textwidth]{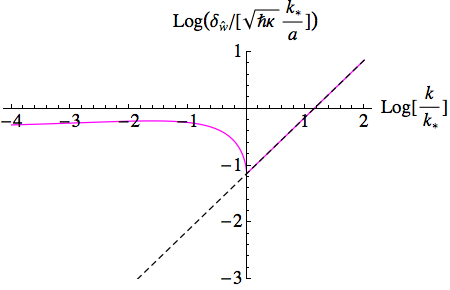}
\hspace{1cm}
\includegraphics[width=0.45\textwidth]{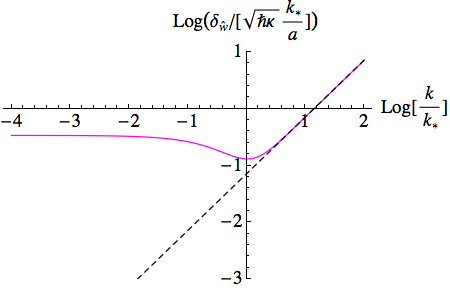}
\includegraphics[width=0.45\textwidth]{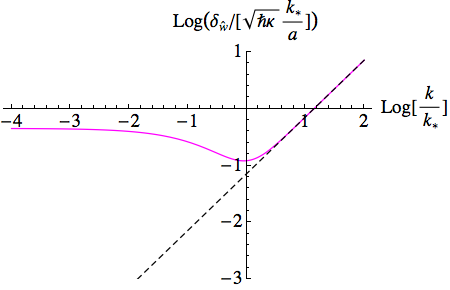}
\caption{Amplification of the primordial gravitational waves by the bounce. The black dashed lines represent the initial amplitude spectrum of the vacuum fluctuations and the magenta solid lines represent the final amplitude spectrum. The final spectrum flattens for wave-numbers close to $\mathbf{k}=\mathbf{k}_*$, where it takes a value dependent on $\mathbf{k}_*/a$. {\it Top left:} The result of the thin-horizon approximation. {\it Top right:} The result of the analytical approximation. {\it Bottom:} The result of the numerical computations.}
\label{fig2}
\end{figure}

In Fig. \ref{fig2} we present (the logarithm of) the spectrum computed by three methods: thin-horizon, analytical and numerical. One can see that around $\frac{\mathbf{k}}{~\mathbf{k}_*}=1$ a clearly better approximation to the exact (numerically computed) spectrum is given by the analytical approximation. On the other hand, for $\frac{\mathbf{k}}{~\mathbf{k}_*}$ less than unity by two or three orders of magnitude the spectrum flattens and the thin-horizon approximation gives a slightly better result there. 

More specifically, from the {\it thin-horizon approximation} we find that the spectrum for $\eta>\eta_+$ reads:
\begin{eqnarray}
&\delta_{\widehat{w}}(k)=\dfrac{\sqrt{2z^2\arctan^2\sqrt{z-1}+4z\sqrt{z-1}+2z-1}}{\pi z}~\cdot\left[\sqrt{\hbar\kappa}\dfrac{~\mathbf{k}_*}{~V^{1/3}}\right] ,\end{eqnarray}
and satisfies the following limits:
\begin{equation}
\delta_{\widehat{w}}(k)\big|_{z=1} = \frac{1}{\pi}~\cdot\left[\sqrt{\hbar\kappa}\frac{~\mathbf{k}_*}{~V^{1/3}}\right],~~\delta_{\widehat{w}}(k)\big|_{z\gg 1}\approx  \frac{1}{\sqrt{2}}~\cdot\left[\sqrt{\hbar\kappa}\frac{~\mathbf{k}_*}{~V^{1/3}}\right],
\end{equation}

From the {\it analytic approximation}  we find that the spectrum for $\eta>\eta_+$ reads:
\begin{eqnarray}
&\delta_{\widehat{w}}(k)=\dfrac{\sqrt{1+2\frac{\cosh^2 \pi \chi}{\sinh^2 (\dfrac{\pi^2}{4z})}}}{\pi z}~\cdot\left[\sqrt{\hbar\kappa}\frac{~\mathbf{k}_*}{~V^{1/3}}\right],\end{eqnarray}
where $\chi=\Im(s)$ (see Def. above Eq. (\ref{DEFs})), and satisfies the following limits:
\begin{equation}\label{limitdelta}
\delta_{\widehat{w}}(k)\big|_{z=1} \approx \frac{1,3}{\pi}~\cdot\left[\sqrt{\hbar\kappa}\frac{~\mathbf{k}_*}{~V^{1/3}}\right],~~\delta_{\widehat{w}}(k)\big|_{z\gg 1}\approx \frac{2}{\pi}~\cdot\left[\sqrt{\hbar\kappa}\frac{~\mathbf{k}_*}{~V^{1/3}}\right].
\end{equation}

\subsection{Production of gravitons}
\begin{figure}[!t]
\includegraphics[width=0.6\textwidth]{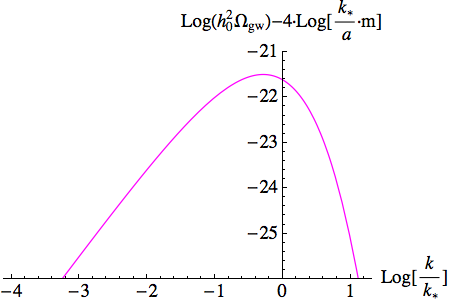}
\caption{Numerical evaluation of the fractional energy of the gravitational radiation per logarithm of frequency $\log ( h_0^2\Omega_{gw})-4\log\left[\frac{\mathbf{k}_{*}}{V^{1/3}}\cdot m\right]$ as a function of $\log\frac{\mathbf{k}~}{\mathbf{k}_*}$. The larger the critical wavenumber $\frac{\mathbf{k}_{*}}{V^{1/3}}$, the larger the energy $\Omega_{gw}$. For $\frac{\mathbf{k}_{*}}{V^{1/3}}=m^{-1}$, we have $4\log\left[\frac{\mathbf{k}_{*}}{V^{1/3}}\cdot m\right]=0$.}
\label{fig_energy}
\end{figure}

We compute for large $\eta_1$ the number of gravitons produced at the bounce. We make use of the formula (\ref{Nv}) which now explicitly reads
\begin{equation}
\langle 0|N_k(\eta_1)|0\rangle=\frac{1}{2}\left(|C_{\mathbf{k}}|^2+|D_{\mathbf{k}}|^2\right)-\frac{1}{2}.
\end{equation}
The energy of gravitons per each polarisation mode and per each wave-vector $\vec{k}:~|\vec{k}|=k$ reads
\begin{align}
E_{k}=\langle 0|N_k(\eta_1)|0\rangle\frac{\hbar ~\mathbf{k}}{V^{1/3}}.
\end{align}
Hence, the energy density of the primordial gravitational waves can be computed as follows:
\begin{align}
\rho_{gw}=\frac{1}{V}\sum_{\vec{k}}E_{|\vec{k}|}\approx\frac{4\pi\hbar}{V^{4/3}}\frac{1}{(2\pi)^3}\int_0^{\infty} \langle 0|N_k(\eta_1)|0\rangle \mathbf{k} k^2\ud k,
\end{align}
which gives a good estimate for $k\gg 1$. In the thin-horizon approximation which gives a good result for $\frac{\mathbf{k}}{~\mathbf{k}_*}\ll 1$, the energy density is estimated to be (again we set $\frak{L}_2=1=\frak{L}_3$, see the discussion of the next section for justification)
\begin{align}
\rho_{gw}\approx\frac{4\pi\hbar}{V^{4/3}}\frac{1}{(2\pi)^3}\int_0^{\mathbf{k}_{*}} \frac{\pi^2}{2}\left(\frac{\mathbf{k}_{*}}{\mathbf{k}}\right)^2 \mathbf{k}^3\ud\mathbf{k}=\frac{\hbar}{8}\left(\frac{\mathbf{k}_{*}}{V^{1/3}}\right)^4,
\end{align}
where we integrate over a finite range of excited wavelengths. A common characterisation of the background of gravitational waves is given by the dimensionless quantity (see e.g. \cite{maggiore}),
\begin{align}\label{dens}
h_0^2\Omega_{gw}(\mathbf{k})=\frac{\kappa}{3H^2_{ref}}~\mathbf{k}\frac{\ud \rho_{gw}}{\ud \mathbf{k}}\approx\frac{z^{-2}}{12}~\cdot\left[\frac{\kappa\hbar}{H^2_{ref}}\left(\frac{\mathbf{k}_{*}}{V^{1/3}}\right)^4\right],~~\mathbf{k}<\mathbf{k}_{*},
\end{align}
where $H_{ref}=100~km/s\cdot Mpc$ and $h_0H_{ref}=H_0$ is the present value of the Hubble rate. $\Omega_{gw}$ is the fractional contribution of the energy of gravitational waves to the total energy of the universe per unit logarithmic frequency. Therefore, we obtain
\begin{align}\label{BEDS}
\log ( h_0^2\Omega_{gw})(\mathbf{k})\approx -17.4+2\log \frac{\mathbf{k}~}{\mathbf{k}_*}+4\log\left[\frac{\mathbf{k}_{*}}{V^{1/3}}\cdot m\right],~~\mathbf{k}<\mathbf{k}_{*}.
\end{align}
The numerical evaluation of the fractional energy of the gravitational radiation per logarithm of frequency $\log ( h_0^2\Omega_{gw})$ is presented in Fig. \ref{fig_energy}. The quantity $4\log\left[\frac{\mathbf{k}_{*}}{V^{1/3}}\cdot m\right]$ should be determined from measurements.

\section{Incorporation of matter-domination epoch}

Since the redshift $z_{eq}=3600$ the evolution of the Universe had been matter-dominated until recently. The matter domination epoch has affected the propagation of gravity waves and therefore, we need to take its effect into account to be sure to obtain a trustworthy estimate of their amplitude spectrum today. We will apply a simplified model of instantaneous radiation-to-matter transition at $z_{eq}$, when $H_{eq}= H_0z_{eq}^{3/2}$ and $V_{eq}= V_0z_{eq}^{-3}$ ($H_0$ and $V_0$ are respectively the present-day values of the Hubble rate and the volume). From this we obtain that for $\eta>0$ in the matter domination epoch
$$a(\eta)=a_0\left(\eta+k_{eq}^{-1}\right)^2,~~a_0=\frac{V_0H_0^2}{4},~~k_{eq}^{-1}=\frac{2}{\sqrt{z_{eq}}~H_0V_0^{1/3}},$$
where $\eta=0$ corresponds to the transition time and the present-day state of the universe is reached at $\eta=(\sqrt{z_{eq}}-1)k_{eq}^{-1}$.

\begin{figure}[!t]
\includegraphics[width=0.55\textwidth]{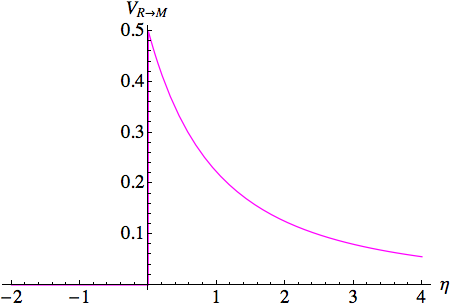}
\caption{The dynamics of the curvature horizon $V_{R\rightarrow M}=\frac{a''}{a}$ in the simplified model of instantaneous  radiation-to-matter transition for $k_{eq}=1/2$.}
\label{fig_transition}
\end{figure}

We study the modes of perturbation away from the bounce, that is, we assume that the redshift $z_{cross}$ at which they entered the curvature horizon after the bounce is much larger than the radiation-to-matter transition redshift, $z_{cross}\gg z_{eq}$. Combining the formulas for the horizon crossing (reentering) time (\ref{hcross}) and for the scale factor (\ref{aR}) we obtain that
\begin{align}
z_{cross}(\mathbf{k})\approx\frac{V_0^{1/3}}{V_{cross}^{1/3}}=\frac{\gamma^{4/3}\sqrt{\frac{\mathbf{k}}{V_0^{1/3}}\frac{\mathbf{k}_*}{V_0^{1/3}}}}{H_0}\gg z_{eq},
\end{align}
where $\frac{\mathbf{k}}{V_0^{1/3}}<\frac{\mathbf{k}_*}{V_0^{1/3}}$ and $V_{cross}=V(\eta_+)$. Therefore:
\begin{align}
\frac{\mathbf{k}_*}{V_0^{1/3}}>\sqrt{\frac{\mathbf{k}}{V_0^{1/3}}\frac{\mathbf{k}_*}{V_0^{1/3}}}\gg 1.6\cdot 10^{-1} Mpc^{-1} .
\end{align}
According to our model, the modes first freely propagate in the radiation-domination epoch and next they instantaneously enter the matter-domination epoch. Eq. (\ref{motion}) (or, Eq. (\ref{eom})) reads
\begin{equation}
v_{\pm,\mathbf{k}}''+(\mathbf{k}^2-V_{R\rightarrow M}(\eta))v_{\pm,\mathbf{k}}=0,
\end{equation}
where
\begin{equation}
V_{R\rightarrow M}(\eta)=\frac{2\theta(\eta)}{(\eta+k_{eq}^{-1})^2}~,
\end{equation}
where $\theta(\eta)$ is the Heaviside function and where we have placed the radiation-matter instantaneous transition at $\eta=0$ (see Fig. \ref{fig_transition}). The solution for $\eta<0$ reads
\begin{equation}\label{wavsol1}
v_{\mathbf{k}}= \frac{\sqrt{4\kappa\frak{L}_2}}{\sqrt{\mathbf{k}}}\left(C_{\mathbf{k}}e^{i{\mathbf{k}}\eta}+D_{\mathbf{k}}e^{-i{\mathbf{k}}\eta}\right),
\end{equation}
whereas for $\eta>0$ it reads 
\begin{equation}\label{wavsol2}
v_{\mathbf{k}}= \frac{\sqrt{4\kappa\frak{L}_2}}{\sqrt{\mathbf{k}}}\left(E_{\mathbf{k}}\bigg(1+\frac{i}{k(\eta+k_{eq}^{-1})}\bigg)e^{i{\mathbf{k}}\eta}+F_{\mathbf{k}}\bigg(1-\frac{i}{k(\eta+k_{eq}^{-1})}\bigg)e^{-i{\mathbf{k}}\eta}\right).
\end{equation}
The values of $C_{\mathbf{k}}$ and $D_{\mathbf{k}}$ have been already determined as the asymptotic values in the respective solution in the bouncing background. The values of $E_{\mathbf{k}}$ and $F_{\mathbf{k}}$ can be determined by sewing the solution (\ref{wavsol1}) with (\ref{wavsol2}) at $\eta=0$. For $\eta>0$, we have (after neglecting the oscillatory term)
\begin{align}\nonumber
&|v_{\mathbf{k}}|^2= \frac{4\kappa\frak{L}_2}{\mathbf{k}}\times\\
&\left[\left(|C_{\mathbf{k}}|^2+|D_{\mathbf{k}}|^2\right)\left(1+\frac{1}{2(\mathbf{k}k_{eq}^{-1})^4}\right)+\frac{1}{(\mathbf{k}k_{eq}^{-1})^4}~\mathit{Re}\left[C_{\mathbf{k}}D^*_{\mathbf{k}}(1+2i\mathbf{k}k_{eq}^{-1}-2(\mathbf{k}k_{eq}^{-1})^2)\right]\right]\\ \nonumber
&\times\left[1+\frac{1}{\mathbf{k}^2(\eta+k_{eq}^{-1})^2}\right],
\end{align}
where $|v_{\mathbf{k}}|^2$ is a function of $\frac{\mathbf{k}}{V_0^{1/3}}$ and $\frac{\mathbf{k}_*}{V_0^{1/3}}$, and  $$\frac{k_{eq}}{V_0^{1/3}}=\frac{H_0\sqrt{z_{eq}}}{2}\approx 1.4\cdot 10^{-2} Mpc^{-1}.$$  
\begin{figure}[!t]
\includegraphics[width=0.85\textwidth]{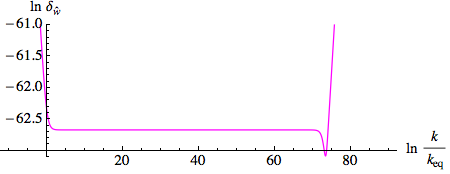}
\caption{The gravity-wave amplitude spectrum for a wide range of wavelengths relative to $\frac{k_{eq}}{V_0^{1/3}}\approx 1.4\cdot 10^{-2} Mpc^{-1}$ for the present-day  with $\frac{\mathbf{k}_*}{V_0^{1/3}}=1.1\cdot 10^{30}Mpc^{-1}$. The amplitude spectrum has three distinct ranges. On the r.h.s., for wave-numbers larger than $\mathbf{k}_*$, the spectrum corresponds to the vacuum state and is linearly growing with $\mathbf{k}$. In the centre, for wave-numbers smaller than $\mathbf{k}_*$ and larger than $k_{eq}$, the spectrum is amplified by the bounce and is flat. On the l.h.s., for wave-numbers smaller than $k_{eq}$, the spectrum is amplified both by the bounce and the radiation-to-matter transition and is growing with decreasing $\mathbf{k}$.}
\label{fig_matter}
\end{figure}
For analysing the final spectrum of the amplitude, it is more convenient to use $\frac{\mathbf{k}_*}{k_{eq}}$ and $\frac{\mathbf{k}_*}{k_{eq}}$ instead of $z=\frac{\mathbf{k}_*}{\mathbf{k}}$ and $\frac{\mathbf{k}}{V_0^{1/3}}$. In Fig. \ref{fig_matter} we plot the amplitude spectrum (\ref{spdef}) for a wide range of scales and for $\frac{\mathbf{k}_*}{V_0^{1/3}}=1.1\cdot 10^{30}Mpc^{-1}$.

\section{Physical implications}
In this section we constrain our simple model by means of the observational bounds on the volume of the present Universe, the redshift at which the bounce occurred and the amplitude of cosmological gravitational wave background. We assume that the universe had been radiation-dominated until $z_{eq}=3600$ and since then it has been matter-dominated until now.

\subsection{Free parameters in quantisation map}
Note that we have constructed a family of singularity resolution models (\ref{hamsemtotal}) that depend on $\frak{L}_1$, $\frak{L}_2$ and $\frak{L}_3$. They are three dimensionless parameters induced by the quantisation map and the semiclassical framework and are parametrised by two numbers, $\mu$ and $\nu$. The value of the repulsive, singularity-resolving potential, $Q_{rep}=\dfrac{\hbar^2\frak{L}_1}{q^2}$, is determined at each $q$ by the value of $\frak{L}_1$. The remaining parameters, $\frak{L}_2$ and $\frak{L}_3$, are dressing the kinetic and potential energy of the gravitational waves in Eq. (\ref{hamsemtotal}) and { they are almost equal to $1$ since, as we show in Subsection \ref{UR}, $\mu,\nu\gtrsim 10^{166}$. Therefore, we are justified in fixing $\frak{L}_2=\frak{L}_3=1$. }

\subsection{Physical solution to the background evolution}
In order to assign to the physical Universe a specific solution of our model we need to determine the values of $(q,p,\frak{L}_1)$ at some moment of the evolution. From the first two quantities we obtain the classical dynamics away from the bounce, whereas the last quantity determines the magnitude of the quantum potential and the early universe dynamics. The last quantity cannot be measured directly but through the amplitude of the gravitational waves, which is encoded into the value of $\frac{\mathbf{k}_{*}}{V^{1/3}}$. 

We shall use another, though equivalent, set of quantities, namely $(V,H,\frac{\mathbf{k}_{*}}{V^{1/3}})$, which are related to the phase space variables as follows (the classical relations):
\begin{align}V=\left(\frac{q}{\gamma} \right)^3,~H=\frac{4\kappa\gamma p}{q^2},~\frac{\mathbf{k}_{*}}{V^{1/3}}=\frac{4\kappa p^2\gamma}{q\sqrt{\frak{L}_1}\hbar}=\frac{H^2V\gamma^2}{4\kappa \sqrt{\frak{L}_1}\hbar}.\end{align}
The wavenumber $\mathbf{k}_{*}$ is dimensionless and its inverse defines the ratio between the shortest amplified wave-length and the length of the universe. Thus,  $\frac{\mathbf{k}_{*}}{V^{1/3}}$ represents the maximal physical wave-number that is amplified. Since the Hubble rate today is known to be
\begin{align}H_0\approx 2.3\cdot 10^{-4} Mpc^{-1},\end{align}
it remains to fix \begin{align}\left(V_0,\frac{\mathbf{k}_{*}}{V_0^{1/3}}\right),\end{align}
in order to determine the full trajectory of the Universe. The present value of the volume is bounded from below:
\begin{align}V_0\gtrsim 3.57\cdot 10^{80} m^3\approx 1.22\cdot 10^{13} Mpc^3.\end{align}
Let us investigate the bounds on $\frac{\mathbf{k}_{*}}{V_0^{1/3}}$.

\subsection{Scale of the bounce}\label{bound1}
According to our model the minimal length of the universe reads
\begin{align}V^{1/3}_{min}=\frac{4\kappa\hbar\sqrt{\frak{L}_1}}{\gamma^2V_0^{2/3}H_0}\sqrt{z_{eq}},\end{align}
(where the above formula takes into account the radiation-to-matter transition). Hence, the bounce took place at the redshift
\begin{align}z_b\approx\frac{V^{1/3}_0}{V^{1/3}_{min}}=\frac{\gamma^2V_0H_0}{4\kappa\hbar\sqrt{\frak{L}_1}}(\sqrt{z_{eq}})^{-1}=\frac{1}{\sqrt{z_{eq}}~H_0}\frac{\mathbf{k}_{*}}{V_0^{1/3}}.\end{align}
It is reasonable to assume that the bounce had occurred before the quark epoch, 
\begin{align}z_b\gtrsim  10^{12},\end{align}
which roughly translates into 
\begin{align}\frac{\mathbf{k}_{*}}{V_0^{1/3}} \gtrsim 10^{10} Mpc^{-1}.\end{align}
Therefore, we expect that the bounce has excited a lot of modes $\frac{\mathbf{k}}{V_0^{1/3}}$, which belong to the interesting, observable range.

\subsection{CMB bound on the amplitude}
From the CMB data collected by the Planck mission, one can determine the value of the amplitude of the scalar power spectrum $A_S=2.2\cdot 10^{-9}$ at the pivot scale of $k_0=0.05 Mpc^{-1}$ and the bound on the tensor to scalar ratio $r<0.12$ \cite{planck13}. Hence, we find
\begin{align}
\delta^2_{\widehat{w}}(k)\big|_{\mathbf{k}\ll\mathbf{k}_{*}}\approx \frac{\hbar\kappa}{\pi^2}\left(z_{LSS}~\frac{\mathbf{k}_{*}}{V_0^{1/3}}\right)^2\lesssim 10^{-10},\end{align}
where $z_{LSS}\approx 1.1\cdot 10^3$ and hence:
\begin{align}
\frac{\mathbf{k}_{*}}{V_0^{1/3}}\lesssim 10^{50}Mpc^{-1}.
\end{align}

\subsection{LIGO bound on the amplitude}
Another constraint comes from the LIGO data \cite{ligo} for the wavelengths in the astrophysical range. Thus, one does not need to take into account the matter-domination epoch which influences only large cosmological scales. The formula for the background energy density spectrum (\ref{BEDS}) can be rewritten as follows
\begin{align}\nonumber
\log (\Omega_{gw})(\mathbf{k})\approx -17.4+2\log \frac{\mathbf{k}~}{\mathbf{k}_{ref}}+2\log\left[\frac{\mathbf{k}_{*}}{V^{1/3}}\frac{\mathbf{k}_{ref}}{V^{1/3}}\cdot m^2\right]-2\log h_0\\=\log \Omega_2+2\log \frac{\mathbf{k}~}{\mathbf{k}_{ref}},
\end{align}
where $\mathbf{k}<\mathbf{k}_{*}$, $\frac{\mathbf{k}_{ref}}{V_0}=\frac{4\pi}{3}\cdot 10^{-7}m^{-1}$, $h_0=6.8$ and thus
\begin{align}
\log \Omega_2=-31.8+2\log \left[\frac{\mathbf{k}_*~}{V_0^{1/3}}\cdot m\right].
\end{align}
In \cite{ligo} it was estimated that $\log \Omega_2\lesssim -7.2$ and hence
\begin{align}
\frac{\mathbf{k}_*~}{V_0^{1/3}}\lesssim 10^{12.3}m^{-1}\approx 6.2\cdot 10^{34} Mpc^{-1}.
\end{align}

\subsection{Constraints put together} 
We have seen that the strongest upper bound on $\frac{\mathbf{k}_{*}}{V_0^{1/3}}$ is given by the LIGO data whereas the strongest lower bound is given by the assumption on the scale of the bounce. When put together, the bounds read:
\begin{align}10^{10} Mpc^{-1} \lesssim  \frac{\mathbf{k}_{*}}{V_0^{1/3}}\lesssim 10^{34}Mpc^{-1},\end{align}
that is, the spectrum of quantum fluctuations of the amplitude of the gravitational waves ``flattens" for wavelengths larger then $\mathbf{\lambda}_{*}=\frac{2\pi}{\mathbf{k}_{*}}$ such that
\begin{align}10^{-12}m\lesssim\mathbf{\lambda}_{*}V^{1/3}_{0}\lesssim 10^{13}m,\end{align}
which at the bounce belonged to the range 
\begin{align}\label{bbwl}10^{-48}m\lesssim\mathbf{\lambda}_{*}V^{1/3}_{min}\lesssim 10~ m.\end{align}
The Eq. (\ref{bbwl}) informs us about the bounds on the minimal wave-length of the amplified mode (at the moment of the bounce) and clearly its value can be as small as $10^{-13}$ Planck lengths or as large as $10^{36}$ Planck lengths. Therefore, in most of the cases (if we use the logarithmic measure of cases) our model does not involve the sub-Planckian modes. Furthermore, we are not particularly interested in the minimal wave-length mode but in all the excited modes that can have much longer wave-lengths . We note that the modes potentially observable in the CMB would be {\it at least} $10^{-12}m$ long at the time of the bounce.

\subsection{Quantum nature of the present-day universe}\label{UR}
Since the cosmological background is given by a wave-function, all classical observables that characterise it are subject to quantum uncertainties. Let us determine the extent of quantum uncertainties in the present state of the universe.

From the obtained bounds on $\frac{\mathbf{k}_{*}}{V_0^{1/3}}$ we get a lower bound on the parameter $\frak{L}_1$,
\begin{align}\sqrt{\frak{L}_1}=\frac{H_0^2V_0\gamma^2}{4\kappa\hbar}\frac{V_0^{1/3}}{\mathbf{k}_{*}}\approx \frac{10^{104}}{Mpc^4}\frac{V_0}{\frac{\mathbf{k}_{*}}{V_0^{1/3}}}\gtrsim 10^{83}.\end{align}
The value of the parameter $\frak{L}_1(\mu,\nu)$ has two components: firstly, it partially follows form the affine coherent state quantisation of the Hamiltonian (which is parameterised by $\nu$) and secondly, it partially follows from the semiclassical approximation based on the affine coherent states (which is parameterised by $\mu$). The bound can be written out as
\begin{align}
\sqrt{\frak{L}_1}=e^{\frac{3}{4\mu}}\sqrt{\frac{\nu+\mu}{2}+\frac{1}{4}}\gtrsim 10^{83},
\end{align}
{ and has two solutions under the assumption that $\nu$ and $\mu$ are of the same order.} Namely,
\begin{align}
\mu\gtrsim 10^{166}~~\textrm{~or~}~~~\mu\lesssim 10^{-3}.
\end{align}
To each solution there corresponds a specific family of the affine coherent states and the respective quantum uncertainties of the basic observables for the universe. A rough estimate (see Appendix \ref{appC} for details) yields
\begin{align}
\frac{\Delta V}{V}\approx \left(e^{1/2\mu}-1\right)^{3/2}
\end{align}
and
\begin{align}
\frac{\Delta H}{H}\approx  \left(e^{1/2\mu}-1\right)^{1/2}+\frac{4\kappa\hbar}{\gamma^2HV}\sqrt{\frac{\mu}{2}}+\left(e^{1/2\mu}-1\right)^{3/2}
\end{align}
For the present-day Universe, $\dfrac{4\kappa\hbar}{\gamma^2H_0V_0}\lesssim 10^{-124}$. We note that the suppression of the quantum uncertainties of the volume and the expansion rate occurs in the present-day universe if $\mu\gtrsim 10^{166}$ (provided that $\sqrt{\frac{\mu}{2}}\ll \frac{\gamma^2H_0V_0}{4\kappa\hbar}$). Therefore, the choice of $\mu$ in that range is consistent with the classicality of the present-day Universe and the bounds on the amplitude of primordial gravitational waves. Furthermore, notice that  $\dfrac{4\kappa\hbar}{\gamma^2HV}$ is growing on the approach to the bounce and that the uncertainty of the expansion rate becomes large in the vicinity of the bounce while the uncertainty of the volume remains constant and small.

\section{Conclusion}

We constructed a quantum model of gravitational waves in a quantum flat FLRW universe filled with perfect fluid. The singularity of the classical model has been replaced by a smooth quantum bounce. The quantum bounce is shown to amplify the amplitude of the gravitational waves. The mechanism of parametric amplification is completely analogous to the one that works in the inflationary scenario. Not surprisingly, the results are similar too as the spectrum of the initial vacuum state vanishing with distance is turned into an almost flat spectrum for wave-lengths larger than some critical value. This value is equal to the inverse of the curvature at the bounce or, in case of inflation, to the inverse of the Hubble rate during the inflation. Just like in case of inflation, we are unable to fix all the relevant parameters of our model. Nevertheless, we impose some constraints from some known observational bounds. Interestingly, we estimated the quantum uncertainties for global geometrical observables, the volume and the Hubble rate, which constitutes a consistency check for our model. 

An important contribution of our model to alternative scenarios lies in the methods that we introduce in the present paper: the coherent state quantisation of cosmological phase space, the use of the affine symmetry and the application of coherent states for the semiclassical approximation of dynamics. In the future we will apply the present approach to investigate the dynamics of density perturbations. This will complete our model and allow to compare it more fully with the Planck data. The results obtained within other approaches such as the Bohmian trajectories approach \cite{cwi} suggest that it will be possible to fit our model to the observed CMB temperature anisotropies and constrain it further.

\appendix
\section{Affine coherent state quantisation}\label{appa}
According to Eq. (\ref{qmap}), the affine quantisation map reads
\begin{align}
f(q,p)\mapsto A_f:=\int_{\mathbb{R}_+\times\mathbb{R}}\frac{\ud q\ud p}{2\pi c}f(q,p)|q,p\rangle\langle q,p|,
\end{align}
where $|q,p\rangle=U(q,p)|\psi\rangle$, $\langle x|q,p\rangle=\frac{e^{ipx}}{\sqrt{q}}\psi(x/q)\in L^2(\mathbb{R}_+,\ud x)$ and $c=\int_{0}^{\infty}|\psi(x)|^2\frac{\ud x}{x}$. For example, for the observables $q^{\alpha}$ and $p$, we find
\begin{align}
A_{q^{\alpha}}=a_{\alpha}x^{\alpha},~~A_p=-i\partial_x+\frac{b}{x},
\end{align}
where $a_{\alpha}=\frac{1}{c}\int_0^{\infty}|\psi(x)|^2\frac{\ud x}{x^{1+\alpha}}$ and $b=\frac{i}{c}\int_0^{\infty}\bar{\psi}(x)\psi'(x)\ud x$. In order to obtain the canonical commutation rule $[A_q,A_p]=i$, we will choose vectors $\psi(x)$ such that $a_1=1$. 

Now, let us make the explicit computation of $A_{p^2}$, that is, the quantum operator that corresponds to the Hamiltonian for the background variables $(q,p)$ in our model and formally is identical with the Hamiltonian of a free particle on the half-line:
\begin{align}
\begin{split}
\langle x|A_{p^2}|y\rangle&=\int_{\mathbb{R}_+\times\mathbb{R}}\frac{\ud q\ud p}{2\pi c}p^2\langle x|q,p\rangle\langle q,p|y\rangle=\int_{\mathbb{R}_+\times\mathbb{R}}\frac{\ud q\ud p}{2\pi c q}\psi(x/q)p^2e^{ip(x-y)}\bar{\psi}(y/q)\\
&=-\int_{\mathbb{R}_+}\frac{\ud q}{c q}\psi(x/q)\delta_{,xx}(x-y)\bar{\psi}(y/q)\\
&=\int_{\mathbb{R}_+}\frac{\ud q}{c q}\delta(x-y)\bar{\psi}(x/q)\bigg(-\psi_{,xx}(x/q)-2\psi_{,x}(x/q)\partial_x-\psi(x/q)\partial_{xx}\bigg)\\
&=\delta(x-y)\bigg(\frac{K_1}{x^2}+\frac{K_2}{x}\partial_x+K_3\partial_{xx}\bigg),
\end{split}
\end{align}
where 
\begin{align}
\begin{split}
&K_1=\frac{1}{c}\int_0^{\infty}\ud z\bar{\psi}(z)\psi'(z)+\frac{1}{c}\int_0^{\infty}z\ud z|\psi'(z)|^2,\\
&K_2=-\frac{2}{c}\int_0^{\infty}\ud z\bar{\psi}(z)\psi'(z),\\
&K_3=-\frac{1}{c}\int_0^{\infty}\frac{\ud z}{z}|\psi(z)|^2=-1.
\end{split}
\end{align}
For the case of a {\bf real vector} $\bar{\psi}(z)=\psi(z)$ we obtain a significant simplification:
\begin{align}
K_1=\frac{1}{c}\int_0^{\infty}z\ud z|\psi'(z)|^2,~~K_2=0,~~K_3=-1.
\end{align}
Let us consider the following family of normalised vectors,
\begin{align}\label{nufiducial}
\psi_{\nu}(x)=\left(\frac{\nu}{\pi}\right)^{\frac{1}{4}}\frac{1}{\sqrt{x}}\exp\left[-\frac{\nu}{2}\left(\ln x-\frac{3}{4\nu}\right)^2\right],~~\nu>0.
\end{align}
Then $a_{\alpha}=\exp\left(\frac{\alpha(\alpha-1)}{4\nu}\right)$ (and $a_1=1$), $b=0$, $K_1=\frac{\nu}{2}$ and hence
\begin{align}
A_{q^{\alpha}}=e^{\frac{\alpha(\alpha-1)}{4\nu}}x^{\alpha},~~A_p=-i\partial_x,~~A_{p^2}=-\partial^2_{x}+\frac{\nu}{2x^2}.
\end{align}
Note the repulsive potential of $A_{p^2}$ that regularises the singularity $x=0$.

\section{Affine coherent states and phase space portraits}\label{appb}
According to Eq. (\ref{seom}) the phase space portrait of a quantum Hamiltonian $\widehat{H}$ can be obtained from the Hamilton equations with the following Hamiltonian:
\begin{align}
H_{S}=\langle q,p|\widehat{H}|q,p\rangle,
\end{align} 
where $|q,p\rangle=U(q,p)|\psi\rangle$ and $\langle x|q,p\rangle=\frac{e^{ipx}}{\sqrt{q}}\psi(x/q)\in L^2(\mathbb{R}_+,\ud x)$. Let us first compute the expectation values of basic operators,
\begin{align}
\langle q,p|x^{\alpha}|q,p\rangle=\beta_{\alpha}q^{\alpha},~~\langle q,p|-i\partial_x|q,p\rangle=\varepsilon p+\frac{\zeta}{q},
\end{align}
where $\beta_{\alpha}=\int_0^{\infty}|\psi|^2x^{\alpha}\ud x$, $\varepsilon=\int_0^{\infty}|\psi|^2\ud x$ and $\zeta=-i\int_0^{\infty}\bar{\psi}\psi'\ud x$. In order for the expectation values of the basic operators to correspond to the parameters of the affine coherent states $q$ and $p$, we will choose vectors $\psi(x)$ such that $\beta_1=1=\varepsilon$ and $\zeta=0$.

Now, let us make the explicit computation for the quantum Hamiltonian,
\begin{align}
\widehat{H}=-\partial^2_x+\frac{K_1}{x^2},
\end{align}
which is the form of the quantum Hamiltonian for the background dynamics in our model and is identical with the Hamiltonian of a particle on the half-line in a repulsive potential $\propto\frac{1}{x^2}$. We find
\begin{align}
\begin{split}
\langle q,p|\widehat{H}|q,p\rangle=-\langle q,p|\partial^2_x|q,p\rangle+\langle q,p|\frac{K_1}{x^2}|q,p\rangle=L_1p^2+L_2\frac{p}{q}+L_3\frac{1}{q^2}+\frac{L_4K_1}{q^2},
\end{split}
\end{align}
where
\begin{align}
\begin{split}
&L_1=\int_0^{\infty} \ud z|\psi(z)|^2,~~L_2=-2i\int_0^{\infty} \ud z\bar{\psi}(z)\psi'(z),\\
&L_3=\int_0^{\infty} \ud z|\psi'(z)|^2,~~L_4=\int_0^{\infty}\frac{\ud z}{z^2}|\psi(z)|^2.
\end{split}
\end{align}
For the case of a normalised {\bf real vector} $\bar{\psi}(z)=\psi(z)$ we obtain a significant simplification:
\begin{align}
L_1=1,~~L_2=0,~~L_3=\int_0^{\infty} \ud z|\psi'(z)|^2,~~L_4=\int_0^{\infty}\frac{\ud z}{z^2}|\psi(z)|^2.
\end{align}
Let us consider the following family of normalised real vectors (note that it is different from Eq. (\ref{nufiducial})),
\begin{align}\label{fiducial}
\psi_{\mu}(x)=\left(\frac{\mu}{\pi}\right)^{\frac{1}{4}}\frac{1}{\sqrt{x}}\exp\left[-\frac{\mu}{2}\left(\ln x+\frac{1}{4\mu}\right)^2\right],~~\mu>0,
\end{align}
which satisfy $\beta_1=1=\varepsilon$ and $\zeta=0$. Then, $L_3=\frac{1}{4}e^{\frac{3}{2\mu}}(1+2\mu)$, $L_4=e^{\frac{3}{2\mu}}$, which combined with the expression for $K_1$ derived in App. \ref{appa} gives,
\begin{align}
H_S=p^2+\frac{e^{\frac{3}{2\mu}}(\frac{\nu+\mu}{2}+\frac{1}{4})}{q^2},~~\langle q,p|x^{\alpha}|q,p\rangle=q^{\alpha}e^{\frac{\alpha(\alpha-1)}{4\mu}}.
\end{align}

\section{Dispersion of affine coherent states}\label{appC}
Given an operator $\widehat{O}$, the dispersion of the affine coherent state $|q,p\rangle=U(q,p)|\psi\rangle$ reads
\begin{align}
\sigma^2_{\widehat{O}}(q,p)=\langle q,p|\widehat{O}^2|q,p\rangle-\langle q,p|\widehat{O}|q,p\rangle^2.
\end{align}
Let $\widehat{O}=x$ be the position operator (a generator of the affine group), then
\begin{align}
\sigma^2_x(q,p)=q^2\left(X_2-X_1^2\right),
\end{align}
where $X_{\alpha}=\int_0^{\infty}|\psi|^2x^{\alpha}\ud x$. For the vector (\ref{fiducial}) we find
\begin{align}
\sigma^2_x(q,p)=q^2\left(e^{1/2\mu}-1\right).
\end{align}
Let $\widehat{O}=\widehat{D}:=-\frac{i}{2}(x\partial_x+\partial_xx)$ be the dilation operator (a generator of the affine group), then
\begin{align}
\sigma^2_{\widehat{D}}(q,p)=q^2p^2(X_2-X_1^2)+qp(I_2-I_1X_1)+(J_2-\frac{I_1^2+1}{4}),
\end{align}
where $I_{\alpha}=\frac{1}{i}\int_0^{\infty}x^{\alpha}(\bar{\psi}\psi'-\bar{\psi}'\psi)^2\ud x$ and $J_{\alpha}=\int_0^{\infty}x^{\alpha}\left|\psi'\right|^2\ud x$. For the vector (\ref{fiducial}) we find
\begin{align}
\sigma^2_{\widehat{D}}(q,p)=q^2p^2\left(e^{1/2\mu}-1\right)+\frac{\mu}{2}\hbar^2.
\end{align}

{
\section{List of symbols}\label{appE}
\begin{center}
    \begin{tabular}{ | l | c |  p{8cm} |}
    \hline
    Symbol & Definition & Meaning \\ \hline
    $q,p$ & \eqref{defa}, \eqref{defa0} & canonical FLRW background variables \\ \hline
    $\delta q_{ab},~\delta\pi^{ab},~w_{ab},~P^{ab}$ & \eqref{defP0}, \eqref{tensor}, \eqref{defP1} & canonical tensor perturbation variables \\ \hline
    $\widecheck{w}_{\pm},~\widecheck{\pi}^{\pm}$ &  \eqref{defFP1}, \eqref{defFP0} & Fourier modes of canonical perturbation variables with polarisation $+$ and $-$\\ \hline
     $\gamma$ & below \eqref{defa} & dimensionless constant \\ \hline
      $T$ & \eqref{defS0}, \eqref{matcon} & fluid variable in the role of internal clock, see the reduction \eqref{defS1} \\ \hline
      $C$ & \eqref{defC} & integration constant of semiclassical motion of the FLRW background \eqref{bgeq}, the value of the semiclassical zero order Hamiltonian \\ \hline
    $\nu$ & \eqref{nufiducial}, below \eqref{covaff2} & positive value parametrising the families of the affine coherent states used for quantisation \\ \hline
    $\mu$ & \eqref{fiducial}, below \eqref{defSUBSET} & positive value parametrising the families of the affine coherent states used for the Klauder framework \\ \hline
  
    $K_1,\frak{K}_i$ & below \eqref{qham}, Appendix \ref{appa} & dressing constants induced by quantisation \\ \hline
    $\frak{L}_i$ & below \eqref{hamsemtotal}, Appendix \ref{appb} & dressing constants induced by quantisation and the Klauder framework \\ \hline
 
    $\mathcal{S}_{\eta}$ & \eqref{defSeta} & FLRW background-dependent factor that transforms the quantum dynamics of gravity waves from the internal clock $T$ to conformal time $\eta$\\ \hline
    
    $v_{g}$ & \eqref{speedofgravity} & speed of gravity waves \\ \hline
   
    $\mathbf{k}$ & \eqref{defk} & dressed co-moving wavenumber of a gravity wave\\ \hline
    $\mathbf{k}_{*}$ & \eqref{defk} & maximal dressed co-moving wavenumber of the gravity waves that are amplified by the bounce \\ \hline
     $z$ & below \eqref{defz} & ratio between the critical and a given wavenumber  \\ \hline
    $V(\eta)$ & by \eqref{eomV} & potential induced by the bounce, against which gravity waves scatter\\ \hline
   $\delta_{\widehat{w}}(k)$ & \eqref{spdef} & amplitude of gravity waves \\ \hline
   $\Omega_{gw}$ & \eqref{dens} & fractional contribution of the energy of gravitational waves to the total energy of the universe per unit logarithmic frequency \\
    \hline
    \end{tabular}
\end{center}
}

\end{document}